\documentclass[a4paper,12pt]{extarticle}

	\newif\ifcomments
	\commentstrue 

\usepackage{comment}
\usepackage{bm}

\usepackage{mathtools}

\usepackage[utf8x]{inputenc}
\usepackage[english]{babel}
\usepackage{amsmath}
\usepackage{amsthm}
\usepackage{amssymb}
\usepackage[linesnumbered,ruled,noend,noline]{algorithm2e}
\usepackage{float}
\usepackage{mathtools}
\usepackage{cleveref}

\usepackage{graphicx}
\usepackage{comment}
\usepackage{tocloft}
\usepackage{tikz}
\usepackage[explicit]{titlesec}
\usepackage{listings}
\usepackage{pdfpages}
\usepackage{pdflscape}

\usepackage [autostyle, english = american]{csquotes}

\usepackage{xcolor}
\usepackage{soul}

\usepackage[toc,page]{appendix}

\usepackage[numbers,sort&compress,merge]{natbib}
\usepackage{chapterbib}
\usepackage[hyphens]{url}

\usepackage{siunitx}
\usepackage{color}
\usepackage[shortlabels]{enumitem}
\usepackage{mathabx}
\usepackage{thmtools,thm-restate}
\usepackage{multirow}
\usepackage{xspace}

\newtheorem{theorem}{Theorem}[section]
\newtheorem{definition}[theorem]{Definition}

\newtheorem{lemma}[theorem]{Lemma}

\newcommand{\D}{\mathcal{D}}

\newcommand\restr[2]{{
\left.\kern-\nulldelimiterspace 
#1 
\vphantom{\big|} 
\right|_{#2} 
}}

  \def\D{\mathcal{D}}

\usepackage[normalem]{ulem}

\ifcomments 
	\def\commentifenable#1{#1}
\else
	\def\commentifenable#1{}
\fi

\def\Re{{\mathbb{R}}}
\def\D{\mathcal{D}}

\newcommand\omaxc[1]{O^{#1}_{\textit{max}}}
\newcommand\ominc[1]{O^{#1}_{\textit{min}}}

\title{Localization with Single or Antipodal Distance Measurements}
\author{Barak Ugav \and Steven M.~LaValle \and Dan Halperin}

\begin{document}

\date{}
\maketitle

\begin{abstract}
	Given a polygon $W$, a \emph{depth sensor}
	placed at point $p=(x,y)$ inside $W$ and oriented in direction $\theta$
	measures the distance $d=h(x,y,\theta)$ between $p$ and the closest point on the boundary of $W$
	along a ray emanating from $p$ in direction $\theta$.
	We study the following problem: For a polygon $W$ with $n$ vertices, possibly with holes, preprocess it such that given a query real value $d> 0$, one can efficiently compute the preimage $h^{-1}(d) \subset W\times \mathbb{S}^1$, namely
	determine all the possible poses (positions and orientations)
	of a depth sensor placed in $W$ that would yield the reading $d$, in an output-sensitive fashion.
	We describe such a data structure requiring $O((E+n) \log n)$ preprocessing time, $O(E)$ storage space and $O(k \log n)$ query time, where $E$ is the number of edges in the \emph{visibility graph} of $W$ (the graph whose nodes are the vertices of $W$, and a pair of nodes is connected by an edge if and only if the open line segment connecting the corresponding vertices is fully contained in $W$), and $k$ is the number of vertices and maximal arcs of low degree algebraic curves constituting the answer.
	We also obtain analogous results for the more useful case (narrowing down the set of possible poses), where the sensor performs two antipodal
	depth measurements from the same point in $W$.
	The query time in the latter structure is $O(\log n+k)$, while the preprocessing time and storage are the same as in the former structure.
	We then describe simpler data structures for the same two problems, where
	we employ a decomposition of $W\times \mathbb{S}^1$, which is an extension of the well-known trapezoidal decomposition,
	and which we refer to as \emph{rotational trapezoidal decomposition}.
	The answers to queries in these data structures are output-sensitive relative to this decomposition: If $k$ cells of the decomposition contribute to the final result,
	we will report them in $O(k)$ (resp., $(\log n + k)$) time for one measurement (resp., antipodal measurements), after $O((E + n) \log n)$ preprocessing time and using $O(E)$ storage space.
	These problems are inspired by \emph{localization} questions in robotics, where we are given a map
	of the environment, a robot is placed in an unknown pose, and we wish to determine where the robot is.
	Although localization is often carried out by exploring the full \emph{visibility polygon} of
	a sensor placed at a fixed point of the environment,
	the approach that we propose here opens the door to sufficing with only few depth measurements,
	which is advantageous as it allows for usage of
	inexpensive sensors and could also lead to savings in storage and communication costs.
\end{abstract}

\section{Introduction}

A depth sensor that provides a single distance measurement\footnote{Throughout the paper we use \emph{depth measurement} and \emph{distance measurement} interchangeably.} is placed inside a known polygonal workspace~$W \subset \Re^2$ with $n$  vertices.
Consider the depth mapping $d = h(x,y,\theta)$ in which $(x,y) \in W$ is the position of the sensor and $\theta \in \mathbb{S}^1$ (namely $[0,2\pi)$) is the direction of the ray emanating from the sensor and along which the distance to the workspace boundary is measured. The set of all sensor poses $(x,y,\theta)$ forms a three-dimensional configuration space (C-space for short).
We wish to devise a data structure such that, after preprocessing the workspace, will efficiently answer the following queries: Given a distance measurement (real number) $d>0$, report all the possible sensor configurations that could yield such a measurement. In other words, we aim to compute $h$'s preimage $h^{-1}(d) = \{(x,y,\theta) \in W \times \mathbb{S}^1 \;|\; d = h(x,y,\theta)\}$.

Determining a sensor's location is one of the main tasks of sensor fusion systems.
Examples are ubiquitous, including Global Positioning Systems (GPS), celestial navigation, trilateration, camera pose estimation, and localization of autonomous vehicles.
Our problem is motivated mostly by the last one, in which a mobile robot has a complete map of its environment and must use sensing to determine its \emph{pose} (position and orientation) with respect to the map.
This so-called {\em robot localization} problem is crucial to robotics and has been researched for several decades.

\subsection{Related Work}
Many variants of the robot-localization problem were studied throughout the years, both where the environment is known and the robot should localize itself in it, or where the environment is not known in advance
and the robot should simultaneously learn and map the environment, and localize itself.

The simultaneous localization and mapping (SLAM) variant, in which the environment is not known, is harder in the sense that the robot has less information than the variant where the environment is known, and therefore methods aiming to solve it often yield imprecise results (see, for example, \cite{DBLP:journals/ram/Durrant-WhyteB06, DBLP:reference/robo/StachnissLT16}).
The robot is required to move and probe
the environment using its sensor(s) (such as a camera, a distance probe, etc.) multiple times, acquiring sufficient information to identify its surrounding.
Earlier work used find-grained grids (typically square or hexagonal grids) to discretize the environment and represent free or occupied regions
\cite{DBLP:books/sp/90/Elfes90,DBLP:journals/computer/Elfes89}.
A common alternative to the grid discretization, is to represent the map as a collection of sparse features extracted from the sensors data stream \cite{DBLP:journals/trob/DissanayakeNCDC01}.
Since the $1990$s, a variety of probabilistic techniques dominate the field \cite{DBLP:books/daglib/0014221}.
The SLAM approach is unavoidably affected by a great deal of uncertainty, and these probabilistic techniques are crucial in order to be able to overcome this challenge,  as they do not commit to possibly noisy observations, but rather use probability distributions.
Sensors commonly used are GPS, compasses, lasers and radars, which are essential for gathering sufficient data about the environment where the robot moves, and while they provide the information required to understand the robot position, all of them yield measurements with errors in real world use, and the probabilistic techniques deal with the noise of these measurements.
While the robot moves and samples measurements using its sensors, an estimation of the environment and the current position are maintained, using a distribution over the range of possible values.
Examples of such commonly used probabilistic techniques are SLAM solutions using (Extended) Kalman Filter \cite{10.1115/1.3662552, DBLP:books/sp/90/SmithSC90, DBLP:journals/trob/DissanayakeNCDC01, DBLP:journals/trob/GuivantN01},  which is an algorithm that estimates the state of the environment and the robot given a series of measurements observed over time, including statistical noise and other inaccuracies.
The classical Kalman Filter assumes that the dynamics of the system are linear, while the Extended Kalman Filter is a nonlinear version of the algorithm.

As part of a SLAM algorithm, a subproblem arises, of {\em incremental} localization (also called {\em relative}, or {\em position/pose tracking}), in which the configuration estimate was known a moment ago and must be updated based on recent motions and sensor readings; see for example \cite{DBLP:conf/iser/MoutarlierC89}.
In case the robot already mapped the environment, but experiences failures and looses its current pose, another subproblem is encuntered, the so-called {\em kidnapped robot problem} \cite{DBLP:conf/icra/EngelsonM92} (also called {\em absolute}, {\em relocation} or {\em global localization}), in which the robot must determine its configuration ``from scratch''.
In the latter variant, the environment is assumed to be known and we are only required to identify where the robot is within it.
The kidnapped-robot version resembles our problem and could correspond to
such a robot using SLAM, or to a freshly deployed or rebooted robot that has a description of the environment in the first place.

Previous research on the kidnapped-robot problem took different approaches, such as discretizing the environment into grid-like cells \cite{DBLP:conf/aaai/BurgardFHS96, DBLP:conf/icra/ChoiCSC10}.
Then, the robot should be located in one of these cells, for example by maintaining a real value for each cell that represent the probability the robot is located in that cell and updating these values when further measurements are available.
A slightly different approach is to use sample random points, instead of using fixed grids, and estimating for each such random point if it is the current position of the robot \cite{DBLP:conf/icra/DellaertFBT99}.

There are both an {\em active} variant \cite{DBLP:journals/siamcomp/DudekRW98,DBLP:journals/trob/JensfeltK01,doi:10.1137/070682885,DBLP:journals/trob/OKaneL07}, in which the robot must determine where to move next to quickly reduce uncertainty, and a {\em passive} variant, in which the sensor fusion system works with whatever sensing and motion data are available.
Passive approaches mainly address stochastic uncertainties \cite{DBLP:books/sp/01/FoxTBD01} and are typically integrated with mapping to obtain SLAM \cite{DBLP:journals/trob/DissanayakeNCDC01,DBLP:journals/arobots/ThrunBF98}.

This work considers the passive localization case and stands out from previous work in that it uses much less sensor data.
Most robot localization methods are based on imaging sensors, such as cameras or Lidars, which provide a dense collection of measurements \cite{DBLP:books/daglib/0014221}.
From a computational geometry viewpoint, these may be considered as providing visibility polygons (see, e.g., \cite{DBLP:journals/dcg/AronovGTZ02, DBLP:journals/comgeo/BoseLM02, DBLP:journals/siamcomp/GuibasMR97,DBLP:journals/comgeo/ZareiG08}), whereas our sensor analogously provides only a single visibility ray with each measurement.
Localization with only few depth measurements is advantageous over exploring the full visibility polygon in that it allows for using inexpensive sensors and requires less storage space and communication.
Our current work, which focuses on one or two measurements, is a step forward in this direction.
However, it still falls short of giving a full localization answer, which in general requires at least three measurements; see further discussion of this issue in Section~\ref{sec:conclusion}.

\subsection{Contribution and Outline}
\begin{figure}[h] \centering
	\includegraphics[scale=0.27]{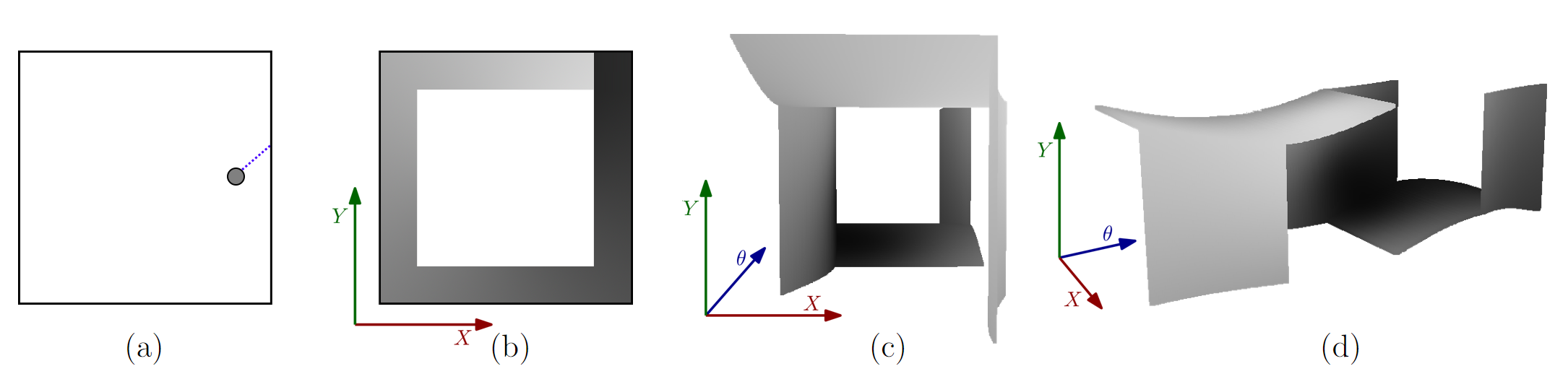}
	\caption{ \label{fig:3d-visual} \sf
		Given a sensor drawn as a gray disc in a unit square workspace (a), and a single distance measurement $d=0.15$ drawn as a dotted line sourced at the sensor and hitting the wall, the possible locations and configurations of the sensor are shown in the two-dimensional workspace (b), and in the three-dimensional C-space (c,d), respectively.
		The gray scale corresponds to the measurement angle: configurations with greater measurement angle are drawn in darker gray.
		The gray shades are projected onto the two-dimensional workspace at (b); note that, as there are multiple configurations with different measurement angles that are projected onto the same position $p$, the shade at $p$ is the darkest shade among all of them.
	}
\end{figure}

Let $W$ be a polygon in the plane, possibly with holes, and having $n$ vertices.
The \emph{visibility graph} of $W$ is the graph whose nodes are the vertices of $W$, and a pair of nodes is connected by an edge if and only if the open line segment connecting the corresponding vertices is fully contained in $W$.
The contributions described in the paper are as follows:
\newline \textbf{(i)}
We review in \Cref{sec:RTD} an extension of the planar trapezoidal decomposition (see, e.g., \cite[Chapter~6]{DBLP:books/lib/BergCKO08}) to the three-dimensional space $W\times \mathbb{S}^1$.
The decomposition, which we refer to as \emph{rotational trapezoidal decomposition} (RTD, for short), has $O(E)$ cells of constant complexity each, and it can be computed in $O((E+n)\log n)$ time, where $E$ is the number of edges in the visibility graph of $W$, which is bounded by $O(n^2)$ in the worst case.
The RTD is easier to construct than the standard three-dimensional vertical decomposition~\cite{DBLP:journals/dcg/BergGH96,hs-18} and it suits the problem at hand better, as the imaginary walls of the RTD are parallel to the measurement direction.
\newline \textbf{(ii)}
In \Cref{sec:single-measurement-rtd} we use this decomposition to build a simple data structure to answer the preimage (see \Cref{fig:3d-visual}) queries---given a real value $d\geq 0$, determine $h^{-1}(d)$---in $O(k)$ time, where $k$ is the number of cells of the RTD that contain part of the answer.
The preprocessing time remains $O((E+n) \log n)$ and the required storage is $O(E)$---this remains true for the following data structures as well.
We analyze (\Cref{subsec:sinlge-result-2d})
the precise shape of the projection of the preimages $h^{-1}(d)$ onto the workspace. These projections could also serve as an answer to a measurement query as they delineate the regions in the plane (positions only) where the sensor could be.
\newline\textbf{(iii)}
The data structure of contribution~(ii) is output sensitive with respect to the (artificial) RTD.
In \Cref{sec:single-measurement-output-sensitive}
we describe a more elaborate data structure, which is output-sensitive in the sense that the query time is proportional, up to a logarithmic factor, to the number of vertices and maximal arcs of low-degree algebraic curves of the output.
\newline\textbf{(iv)}
In \Cref{sec:antipodal-measurements} we discuss the following two-measurement query type:
Given two antipodal depth measurements of a sensor from the same (unknown) point in $W$ in the (unknown) directions $\theta$ and $\theta +\pi$, determine all the poses of the sensor that would yield these measurements.
In \Cref{subsec:antipodal-measurements-rtd} we devise a data structure that answers the antipodal queries in $O(\log n+k)$ time where $k$ is the number of cells of the RTD that contain part of the answer.
Similar to contribution~(ii), the RTD-based data structure for antipodal measurements is output sensitive with respect to the RTD, and in \Cref{subsec:antipodal-output-sensitive} we present
an output-sensitive data structure for antipodal measurements that answer a query in $O(\log n+k)$ time where $k$ is the number of vertices and maximal arcs of low-degree algebraic curves that constitute the output.
\newline\textbf{(v)} We implemented the RTD-based data structures, and we
demonstrate a few results obtained with the implementation. The code is open source and is available for download; see \Cref{sec:implementation}.

Throughout the paper, we assume general position of the input: no three vertices lie on the same line, no three edges intersect at the same point, no two edges are parallel, etc.
(We give up on the general position assumption in \Cref{fig:3d-visual} for simplicity of demonstration.)

\section{Rotational Trapezoidal Decomposition}
\label{sec:RTD}

We will describe a few data structures that are built upon a known decomposition of the three-dimensional C-space $\mathbb{R}^2\times \mathbb{S}^1$,
which we refer to as \textit{rotational trapezoidal decomposition}, or RTD for short.
This decomposition has been previously used in other problems involving three-dimensional configurations spaces involving translation and rotation in the plane; see, e.g., \cite{DBLP:journals/amai/SharirS91}.
Conceptually, the decomposition generalizes the known, planar, vertical decomposition~\cite[Chapter~6]{DBLP:books/lib/BergCKO08} to any direction.
The planar trapezoidal decomposition partitions the workspace $W$ to a set of trapezoids.\footnote{Some of the two-dimensional cells in the decomposition may be triangles; for brevity we will also refer to them in this context as trapezoids.}
A trapezoid is identified by its ceiling and floor (namely top and bottom edges, respectively), and the left and right vertices, which define the left and right artificial vertical edges.
See \Cref{fig:vertical-decomposition} for an illustration.
We remark that there are other possible decompositions of \emph{arrangements} of surfaces in three-dimensional space~\cite[Section~28.3]{hs-18}, but RTD seems most suitable for our purpose.

We can rotate the scene to any orientation and compute all the trapezoids of any angle.
As we let the orientation $\theta$ of the decomposition vary a little, we notice that the trapezoids remain similar (in a sense to be made precise shortly) unless they undergo some combinatorial change.
As the direction of decomposition changes, the trapezoid changes continuously, and we refer to the union of these trapezoids as a three-dimensional cell, or \emph{cell} for short.
Since throughout the change of orientation the ceiling and floor edges, as well as the vertices determining the left and right boundaries of the trapezoid remain the same we use the following notation:
A cell $C$ is identified by its ceiling (top) and floor (bottom) edges,
$e^C_t,e^C_b$, and the left and right vertices, $v^C_l,v^C_r$, which define the left and right artificial edges (parallel to the direction of the decomposition).
The two limiting vertices, $v^C_l,v^C_r$, may be endpoints of $e^C_t$, $e^C_b$, or other reflex vertices of the workspace.

Now, in addition to describing a cell by its floor, ceiling and the two vertices that define the two artificial side edges, we also add the first and last orientations where the trapezoid exists (in the combinatorial sense, namely having the same edges and vertices defining it).
These cells constitute a decomposition of the three-dimensional C-space of the sensor.
With this description, there are at most $O(n^2)$ unique cells, and we can compute each cell's exact boundary as a function of the orientation $\theta$ of the rotated workspace.
More precisely, there are $O(E)$ such unique cells, where $E$ is the number of edges in the visibility graph of $W$, which is bounded by $O(n^2)$.

\subsection{Computing the Rotational Trapezoidal Decomposition}
\begin{figure}[h] \centering
	\includegraphics[page=7, scale=0.9]{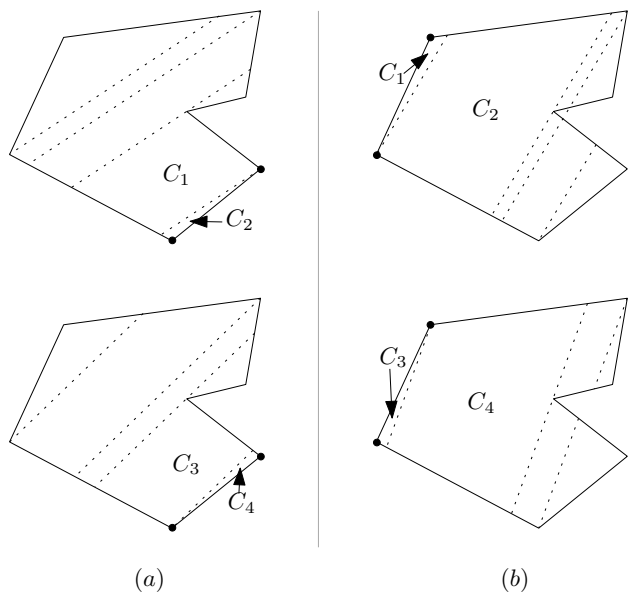}
	\caption{ \label{fig:event-case1} \sf
		Type~I events: Two vertices of the same edge align along the rotating ray.
		Rays are rotating counterclockwise.
		The event vertices are drawn as small black discs.
		Each column depicts one example, before (top) and after (bottom) the event.}
\end{figure}

\begin{figure}[h] \centering
	\includegraphics[page=8, scale=0.85]{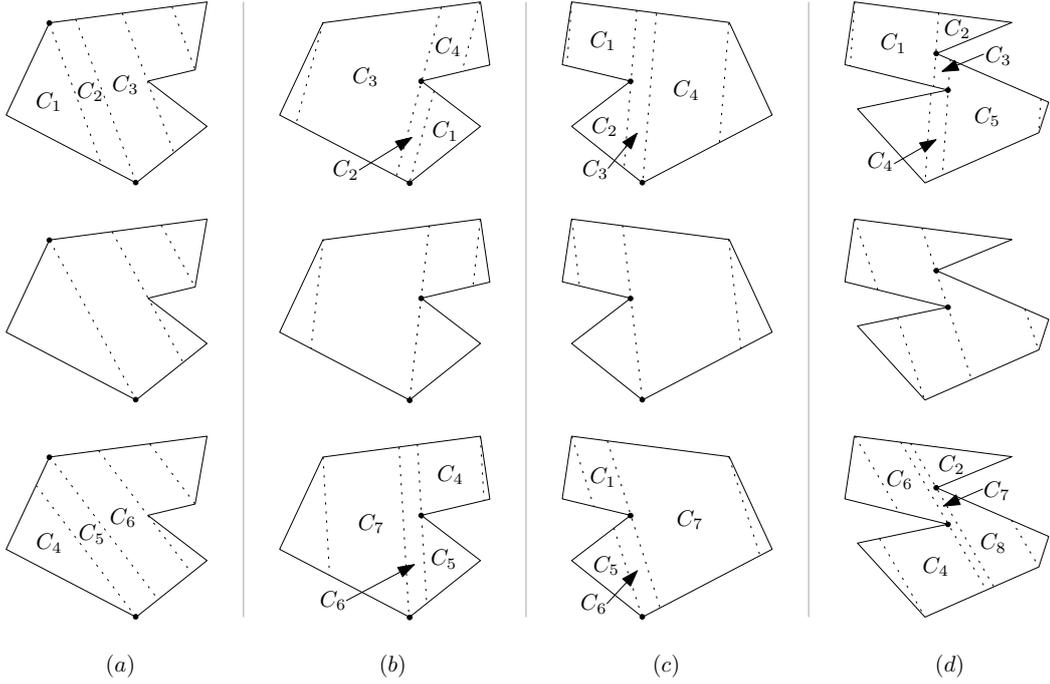}
	\caption{ \label{fig:event-case2} \sf
		Type~II events: Two vertices of two distinct edges align along the rotating ray.
		Rays are rotating counterclockwise.
		The event vertices are drawn as small black discs.
		Each column depicts one example, before (top), during (middle), and after (bottom) the event.}
\end{figure}

We construct the cells of the RTD by simultaneously performing a radial sweep around all the vertices of the workspace, where each vertex is the origin of a sweeping ray, all rays point at the same direction and are rotated together.
An event occurs when one of the rays originating at some vertex $u$ hits another vertex $v$ that is visible from $u$.
We compute all the events in advance by computing the visibility graph (VG) of $W$ in $O(E + n \log n)$ time using the algorithm of Ghosh and Mount~\cite{DBLP:journals/siamcomp/GhoshM91} and creating for each edge $(u,v)$ in the VG two events (one originating at $u$ with direction $\overrightarrow{uv}$ and the other originating at $v$ with the opposite direction), and then radially sorting
all of them by $\theta$.
In addition, each ray stores at all times (angles) the cells on both its sides, allowing us to keep track of all the existing trapezoids of the decomposition at the current orientation.
During the handling of an event, where two vertices align along the sweeping ray (its origin and another vertex),
we know they are visible from one another
(as the events were created from the visibility graph),
namely the open line segment connecting them lies in the interior of the workspace, and cells should be created or terminated.
There are two types of events along the radial sweep:
\begin{description}
	\item[Type~I]
	      Two vertices of the same edge are incident to a rotating ray.
	      See \Cref{fig:event-case1} for an illustration.
	      The triangular cell that contained both vertices on its boundary is ``squeezed'' and terminated, and a new triangular cell is created.
	      The cell sharing an artificial edge with the triangular cell also terminates and a new one is created due to change of one of its limiting vertices.
	      In total, two cells are terminated and two are created.

	\item[Type~II]
	      Two vertices of two different edges are incident to a rotating ray.
	      See \Cref{fig:event-case2} for an illustration.
	      The middle cell is ``squeezed'' and terminated, and a new middle cell is created.
	      The cells from both sides of the connecting ray terminate and new ones are created due to change in the limiting vertices.
	      In total, three cells are terminated and three are newly created.
\end{description}

In both event types, the top and bottom edges and left and right vertices are known locally from the information maintained by the rays or the terminated cells.
During handling of an event we terminate or create a constant number of cells.
As there are $O(E)$ events, there are also $O(E)$ cells in total.

To start the process, trapezoidal decomposition in the horizontal direction (namely, with $\theta=0$), is performed, requiring $O(n\log n)$ time and starting $\Theta(n)$ cells.
We let $\theta$ vary in the range $[0,2\pi)$.
Notice that the trapezoidal decomposition at $\theta=0$ artificially cuts cells into two; this however does not affect the analysis that follows or the asymptotic resources required by the algorithm.
In total, we construct these $O(E)$ cells and their angle intervals in $O((E+n) \log n)$ time, while using $O(E+n)$ space.
It is easy to see that $\Omega(n^2)$ cells may be created and so the bound on the maximum number of cells is tight.
The process is summarized in the following lemma.

\begin{lemma}
	Given a polygonal workspace $W$ with a total of $n$ vertices, the three-dimensional configuration space $W \times \mathbb{S}^1$ is partitioned by the Rotational Trapezoidal Decomposition (RTD) described above into $O(E)$ cells, each of constant descriptive complexity, where $E$ is the number of edges in the visibility graph of $W$.
	The construction of the decomposition takes $O((E+n) \log n)$ time and requires $O(E+n)$ storage space.
	The number of edges $E$ can be $\Theta(n^2)$ in the worst case.
\end{lemma}

\subsection{The RTD Opening Function}
\label{subsec:rtd-opening-func}

Each cell $C$ prevails through an interval of $\theta$ values, calculated during the sweep---we denote this interval by $\Theta^C=[\theta^C_{\textit{begin}},\theta^C_{\textit{end}})$.
Given a cell $C$ and an angle $\theta \in \Theta^C$, we can easily construct the trapezoid that is the cross-section of $C$ at $\theta$.
The top left and top right vertices (which are not necessarily vertices of the workspace) of the trapezoid both lie on $e^C_t$.
We can compute their $x$ values as a function of $\theta$, denoted $x^C_{tl}(\theta),x^C_{tr}(\theta)$.
Assume, w.l.o.g., that the top and bottom edges are not vertical, and let $y=e^C_t(x)$ be the function that maps an $x$-coordinate to a $y$ value of a point $(x,y)$ on $e^C_t$, and similarly $y=e^C_b(x)$ for the bottom edge.

\begin{definition} \label{def:opening-func}
	The \emph{opening of a cell} $C$ at angle $\theta$ and at a point $(x, e^C_t(x))$ on the top edge of the cell, is the length of the intersection of the line with slope $\theta$ through the point $(x,e^C_t(x))$ on the top edge of the cell, with the trapezoidal cross-section of $C$ at angle~$\theta$. We denote this opening by $O^C(\theta,x)$.
\end{definition}

Next, we show a few properties of the opening function:
\begin{itemize}
	\item The opening function can be analytically described as a function of $\theta$ and $x$:
	      Given a cell $C(e^C_t,e^C_b,v^C_l,v^C_r,\theta^C)$, assume w.l.o.g that the top and bottom edges are not vertical, let $e^C_t(x)=x m_{e_t} + b_{e_t}$ be the equation for the top edge and analogously for the bottom edge $e^C_b(x)=x m_{e_b} + b_{e_b}$.

	      \begin{figure}[h] \centering
		      \includegraphics[page=1, scale=1]{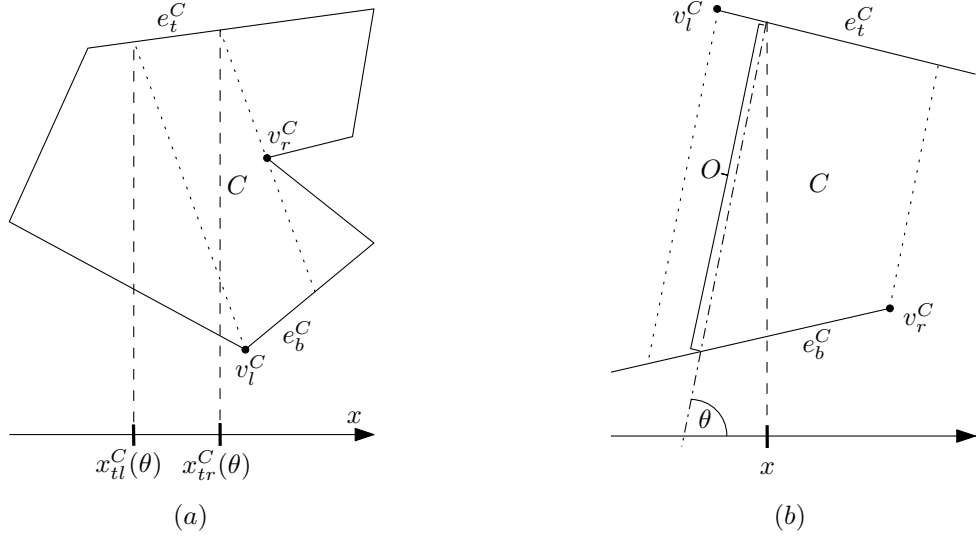}
		      \caption{\label{fig:opening-func} \sf
			      (a) The $x$ values of the left and the right vertices of a cell's ceiling.
			      (b) The opening size $O$ as a function of $x$ and $\theta$.}
	      \end{figure}

	      \begin{figure}[h] \centering
		      \includegraphics[scale=0.5]{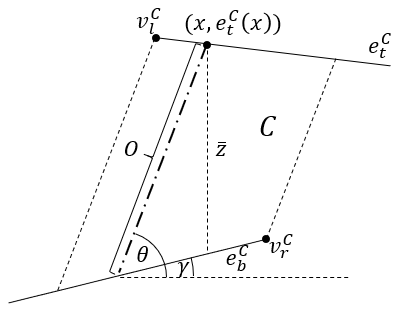}
		      \caption{\label{fig:opening-calc} \sf The opening function of a single cell, for fixed values of $x$ and $\theta$, is drawn along with the vertical segment $\bar{z}$ at $x$.}
	      \end{figure}

	      For any $x$, denote by $\bar{z}$ the vertical segment at $x$ with one endpoint on $e^C_b$ and the other endpoint on $e^C_t$, and denote its length by $|\bar{z}|$.
	      Consider the triangle whose edges are $O,\bar{z}$ and a portion of $e^C_b$ (see \Cref{fig:opening-calc} for an illustration).
	      The angle opposite $e^C_b$ is $\frac{\pi}{2} -\theta$, the angle opposite of $\bar{z}$ is $\theta -\gamma$ and the angle opposite $O$ is $\frac{\pi}{2} +\gamma$.
	      Using the law of sines:

	      \[ O(\theta,x) =\frac{|\bar{z}| \sin (\frac{\pi}{2} + \gamma)}{\sin (\theta -\gamma)} =\frac{ \cos \gamma}{\sin (\theta - \gamma)} =\frac{ \cos \gamma}{\sin \theta \cos \gamma - \cos \theta \sin \gamma} \]
	      \[ =\frac{ \frac{1}{\sqrt{1+m^2 _{e_b}}}}{\sin \theta \frac{1}{\sqrt{1+m^2 _{e_b}}} - \frac{m_{e_b}}{\sqrt{1+m^2 _{e_b}}} \cos \theta} =\frac{}{\sin \theta - m_{e_b} \cos \theta} =\frac{e^C_t(x)-e^C_b(x)} {\sin \theta - m_{e_b} \cos \theta} \]
	      \[ =\frac{ x(m_{e_t} - m_{e_b}) + b_{e_t} - b_{e_b} } {\sin \theta - m_{e_b} \cos \theta} \;. \]

	\item The inverse of the opening function can be analytically described:
	      For each RTD cell $C$, solving $O(\theta,x)=f$ for a specific value $f$ is required by later data structures.
	      Given a cell $C(e^C_t,e^C_b,v^C_l,v^C_r,\theta^C)$, assume w.l.o.g that the top and bottom edges are not vertical, and denote by $e^C_t(x)=x m_{e_t} + b_{e_t}$ the equation for the top edge and equivalently for the bottom edge $e^C_b(x)=x m_{e_b} + b_{e_b}$.
	      Using the opening equation:
	      \[ O^C(\theta,x)=f \;, \]
	      \[ \frac{e^C_t(x)-e^C_b(x)} {\sin \theta - m_{e_b} \cos \theta}=f \;, \]
	      \[ e^C_t(x)-e^C_b(x))= f(\sin \theta - m_{e_b} \cos \theta) \;, \]
	      \[ m_{e_t}x +b_{e_t}-m_{e_b}x -b_{e_b}= f(\sin \theta - m_{e_b} \cos \theta) \;, \]
	      \[ x=\frac{f(\sin \theta - m_{e_b} \cos \theta) + b_{e_b} - b_{e_t}}{m_{e_t} -m_{e_b}} \;. \]

\end{itemize}

\subsection{The Vertices of a Trapezoid}

For each RTD cell $C$, which exists in an angle interval $\Theta^C$, an exact trapezoid description can be calculated analytically for any fixed $\theta\in\Theta^C$.
In this section we give an exact formula for the two top vertices of a trapezoid, $x^C_{tl}(\theta),x^C_{tr}(\theta)$.

The top and bottom edges are fixed, and the two artificial edges are oriented at angle $\theta$.
It remains to express the vertices created by the intersections of the artificial edges and the top edge, denoted $v_{tl},v_{tr}$ with $x$ values $x^C_{tl}(\theta),x^C_{tr}(\theta)$.
Given a cell $C(e^C_t,e^C_b,v^C_l,v^C_r,\theta^C)$, assume, w.l.o.g., that the top and bottom edges are not vertical, and denote by $e^C_t(x)=x m_{e_t} + b_{e_t}$ the equation for the top edge and similarly for the bottom edge $e^C_b(x)=x m_{e_b} + b_{e_b}$.
Let $(x_{v_{tl}},y_{v_{tl}})$ be the coordinates of the left vertex $v^C_l$ defining the trapezoid, then any point $(x,y)$ on the left artificial edge can be expressed as (see \Cref{fig:opening-func}a):
\[ y=\tan \theta (x - x_{v_{tl}}) + y_{v_{tl}} \;. \]
To calculate the intersection with the top edge, we substitute the top edge equation:
\[ m_{e_t}x+b_{e_t}=x \tan \theta - x_{v_{tl}} \tan \theta + y_{v_{tl}} \;, \]
\[ x(m_{e_t} - \tan \theta)= y_{v_{tl}} - x_{v_{tl}} \tan \theta -b_{e_t} \;, \]
\[ x= \frac{y_{v_{tl}} - x_{v_{tl}} \tan \theta -b_{e_t}}{m_{e_t} - \tan \theta} \;. \]
Similar calculation gives the right top vertex.
In conclusion:
\[
	x^C_{tl}(\theta)=\frac{y_{v_l} - x_{v_l} \tan \theta -b_{e_t}}{m_{e_t}- \tan \theta}
	\;,\quad
	x^C_{tr}(\theta)=\frac{y_{v_r} - x_{v_r} \tan \theta -b_{e_t}}{m_{e_t}- \tan \theta}
	\;.
\]

\section{RTD-Based Single Distance Measurement Localization}
\label{sec:single-measurement-rtd}

In this section we describe a data structure to determine the possible configurations of a sensor given a single distance measurement $d$ taken from an unknown position in an unknown direction.
The data structure uses RTD, the decomposition described in \Cref{sec:RTD}, and achieves a query time of $O(k)$, where $k$ is the number of RTD cells that contains part of the result.
Note that such query time is optimal with respect to the RTD, but fails to achieve real output sensitivity---this requires a more involved data structure, which we will present in the next section, \Cref{sec:single-measurement-output-sensitive}.

\subsection{RTD-Based Data Structure Construction}
We use the decomposition described in \Cref{sec:RTD} of the three-dimensional C-space of the sensor into cells of constant descriptive complexity each, such that within each cell $C$, the sub-region $C_d=\{(x,y,\theta) \in C \;|\; d = h(x,y,\theta)\}$ has constant descriptive complexity as well.
We then show, how to maintain the cells of the decomposition in a search structure, such that given a query $d$ we can easily retrieve all the cells $C$ for which $C_d\neq \emptyset$, in time proportional to their number.

\begin{figure}[h] \centering
	\includegraphics[page=3, scale=0.65]{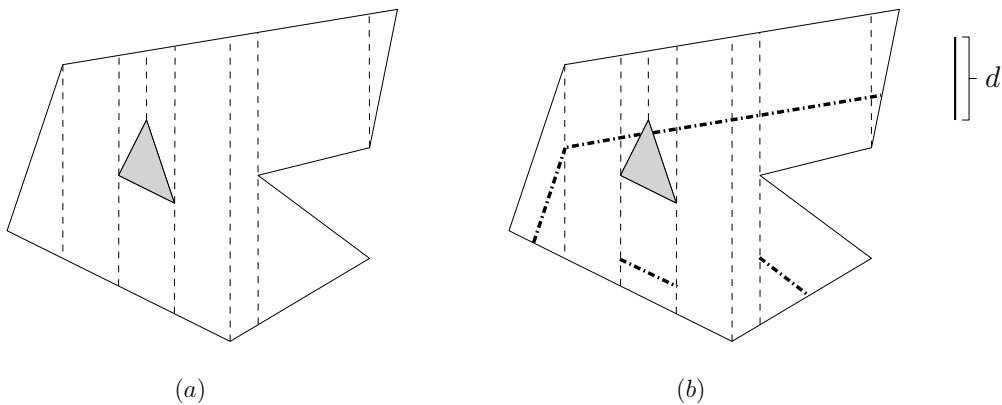}
	\caption{ \label{fig:vertical-decomposition} \sf
		Example of a vertical decomposition of a polygonal workspace with a single hole (a) and the line segments (bold, dash-dotted) along which the sensor may be when it reads the distance $d$ in the upward vertical direction (b).}
\end{figure}

Suppose for the purpose of exposition that we know that the sensor measures its distance in the upward vertical direction, namely $\theta=\pi/2$.
We apply trapezoidal (vertical) decomposition to the workspace.
Given a measurement $d$, we can easily compute for each trapezoid in the decomposition whether there are any points in it that are at vertical distance $d$ from the top edge of the trapezoid.
These points lie along a segment parallel to the top edge, and the desired answer is the union all such segments over all trapezoids.
See \Cref{fig:vertical-decomposition} for an illustration.

The direction of trapezoidal decomposition is parallel to the direction of measurement, therefore for us to remove the assumption that the sensor measures its distance in the vertical direction, instead of considering the result segments within each trapezoid of the vertical decomposition, we consider the sub-region $C_d$ within each cell $C$ in the RTD decomposition.
Each such cell has constant complexity and the result sub-region is easy to compute.

We wish to store the cells of the decomposition of the C-space of the sensor in a data structure such that given a distance measurement $d$, we can efficiently retrieve all the cells that have a non-empty set of candidate poses for this measurement.
To this end we define the maximal opening $\omaxc{C}$ of a cell $C$, which is the maximal-length $\theta$-oriented segment that can be placed in a $\theta$-cross-section of a cell $C$, over all possible  values $\theta\in\Theta^C$.

We introduced the opening function in the RTD section (\Cref{def:opening-func}). Now we define for a cell $C$ the \textit{maximal opening}, which is the maximum value of the opening function for any valid values of $\theta,x$:
\[ \omaxc{C}=\max_{\theta\in\Theta^C} {\max_{x\in [x^{C}_{tl}(\theta),x^{C}_{tr}(\theta)]} {O^{C}(\theta,x)}} \;. \]

In a preprocessing phase, we calculate $\omaxc{C}$ for each cell,
and keep the cells sorted in decreasing order by this value in an array.
Given a query measurement $d$, the cells that contain potential poses for this reading are exactly the cells for which $d \leq \omaxc{C}$.
We can find these cells by traversing the sorted cells in descending maximum opening order, resulting in $O(k)$ query time,\footnote{For $k=0$ we still need to check the cell at the top of the list to find out that no relevant cells exist.} where $k$ is the number of cells that contain workspace points of the answer.
The calculation of the maximum opening takes constant time per cell.
Since there are $O(E)$ cells overall, the entire preprocessing requires $O((E+n) \log n)$ time and $O(E+n)$ space, which is asymptotically subsumed by the construction of the decomposition (\Cref{sec:RTD}).

\begin{theorem}
	Given a polygonal workspace with a total of $n$ vertices, the RTD cells of the configuration space $W \times \mathbb{S}^1$ can be computed and sorted in decreasing order of their maximum opening in time $O((E+n) \log n)$ and $O(E+n)$ space, where $E$ is the number of edges in the visibility graph of $W$.
	Then, given a query real parameter $d>0$, we can report all the cells that contain a possible configuration yielding depth measurement $d$ in time $O(k)$, where $k$ is the number of the relevant cells.
\end{theorem}

This structure has a couple of advantages: (i) it is trivial to construct and implement (see \Cref{sec:implementation}),
and (ii) it is worst-case optimal in storage space and query time \emph{assuming the Rotational Trapezoidal Decomposition}.
However, RTD subdivides the C-space in a way that may induce large unnecessary overhead for certain queries (see \Cref{subsec:single-measurement-rtd-suboptimality}).
In \Cref{sec:single-measurement-output-sensitive} we provide an improved solution with only a logarithmic factor in query time over the minimum possible output size.

\subsection{Sensor Positions in a Single RTD Cell}
\label{subsec:sinlge-result-2d}

Given a single sensor reading $d$, we concern ourselves in this section with the shape and complexity of the locus of sensor \emph{positions} that result in such a reading, namely ignoring the orientation of the sensor's ray.
We believe that reporting the possible positions only is, although incomplete, useful feedback that is more intuitive and easy to grasp by the user of a robot system.
This in itself can serve as an answer to a query, in particular, when we wish to further process the regions using additional information about the potential sensor pose.

For each three-dimensional cell $C$ of the RTD, let $\Psi_d(C)$ denote the union of  positions where the sensor could be, namely $\Psi_d(C)$ is the projection onto the $xy$-plane of $C_d$.
Fix a cell $C$ with top and bottom edges $e^C_t,e^C_b$, limiting vertices $v^C_l,v^C_r$, which exist in an angle interval $\Theta^C$.
Recall that $x^{C}_{tl}(\theta),x^{C}_{tr}(\theta)$ denote the $x$ value of the left and right top vertices of the trapezoidal cross-section of $C$ (which are not necessarily vertices of the workspace), respectively, at angle $\theta$.

\begin{figure}[h] \centering
	\includegraphics[page=1, scale=0.9]{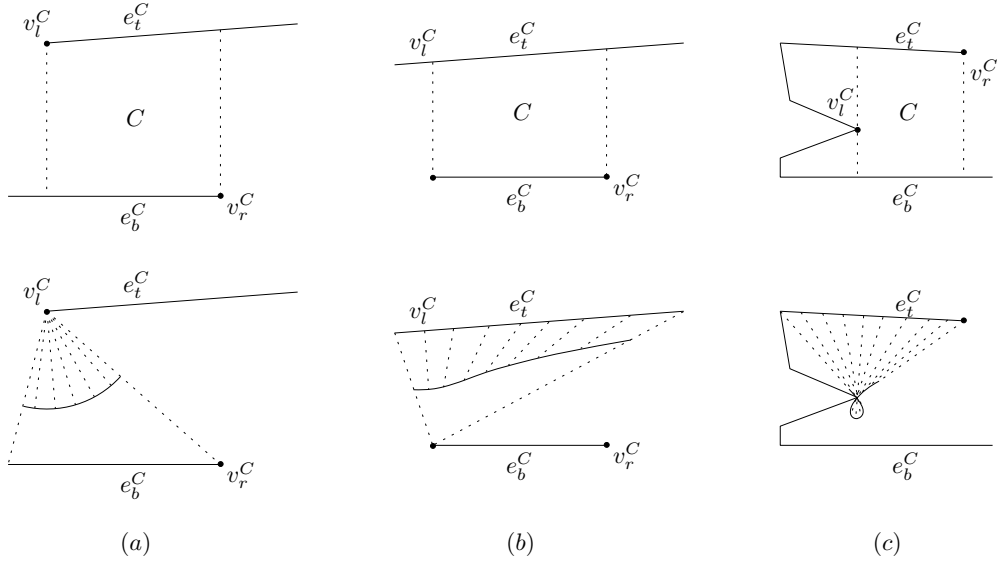}
	\caption{ \label{fig:curveTypes} \sf (a)
		Type~A curve: when the limiting vertex lies on the cell's ceiling, the curve is a circular arc.
		(b)(c) Type~B curve: when the limiting vertex does not lie on the cell's ceiling, the curve is a conchoid of Nicomedes.}
\end{figure}

Let $p^{C}_{l}(\theta)$ denote the position of the sensor, where a distance measure $d$ at angle $\theta$ will hit $e^{C}_{t}$ at $x$ value $x^{C}_{tl}(\theta)$, while ignoring other obstacles in the workspace.
Specifically, the point whose $x$-coordinate is $x^{C}_{tl} (\theta) -d\cos \theta$ and $y$-coordinate is $e^{C}_{t}(x^{C}_{tl} (\theta)) -d\sin \theta$, where $e^{C}_{t}(\cdot)$ is the equation of the line supporting the top edge.
Similarly denote by $p^{C}_{r}(\theta)$ the sensor pose that will result in hitting $e^{C}_{t}$ at $x^{C}_{tr}(\theta)$.
Recall that $\Psi_d(C)$ denotes the region $\{(x,y)|(x,y,\theta)\in C \ {\rm and}\ h(x,y,\theta)=d\}$, namely the positions of the sensor in the workspace, where the sensor configuration is in $C$ and there is a sensor reading $d$.
We wish to trace the curves drawn by $p^{C}_{l}(\theta)$ and $p^{C}_{r}(\theta)$ as $\theta$ varies in the range $\Theta^C$.
These curves are portions of the boundary of the region $\Psi_d(C)$.
The type of the curve that the function $p^{C}_{l}(\theta)$ (respectively, $p^{C}_{r}(\theta)$) draws, depends on $v^{C}_{l}$ (respectively, $v^{C}_{r}$), the vertex defining the left (respectively, right) wall of the cell.
We describe it here for $v^{C}_{l}$.
The description for $v^{C}_{r}$ is analogous.
\begin{description}
	\item[Type~A]
	      The vertex $v^{C}_{l}$ defining the left wall of the trapezoid/cell lies on $e^{C}_{t}$.
	      See \Cref{fig:curveTypes}(a) for an illustration.
	      The top left vertex, and therefore $x^{C}_{tl}(\theta)$, are fixed (i.e., independent of $\theta$), and the curve is a circular arc of radius $d$ around $v^{C}_{l}$: $v^{C}_{l} -(d\cos \theta, d\sin \theta)$.

	\item[Type~B]
	      The vertex $v^{C}_{l}$ defining the left wall of the trapezoid/cell does \emph{not} lie on $e^{C}_{t}$.
	      See \Cref{fig:curveTypes}(b,c) for illustrations.
	      The curve traced by $p^{C}_{l}(\theta)$ is a conchoid of Nicomedes; see \Cref{app:single-projection} for details.
\end{description}

For any angle $\theta \in \Theta^C$, all points on the line segment $(p^{C}_{l}(\theta), p^{C}_{r}(\theta))$ are points the robot can be at and measure $d$ at $e^{C}_{t}$ (ignoring other obstacles).
Together with the types of the curves traced by $p^{C}_{l}(\theta), p^{C}_{r}(\theta)$, we have an exact description of the region of the workspace containing all the points the robot might be at, while its configuration is in the cell $C$.

If the interval $\Theta^C$ contains the angle perpendicular to the top edge, denoted $\theta^C_p$, we divide it into two sub-intervals, one with angles that are greater than or equal to the perpendicular angle and the other with smaller angles, $[\theta^C_{\textit{begin}},\theta^C_p),[\theta^C_p,\theta^C_{\textit{end}})$.
A single shape per sub-interval will be created.
If $\theta^C_p$ is not included in $\Theta^C$, a single interval $[\theta^C_{\textit{begin}}, \theta^C_{\textit{end}})$ and a single corresponding shape is used.

\begin{figure}[h] \centering
	\includegraphics[scale=0.27]{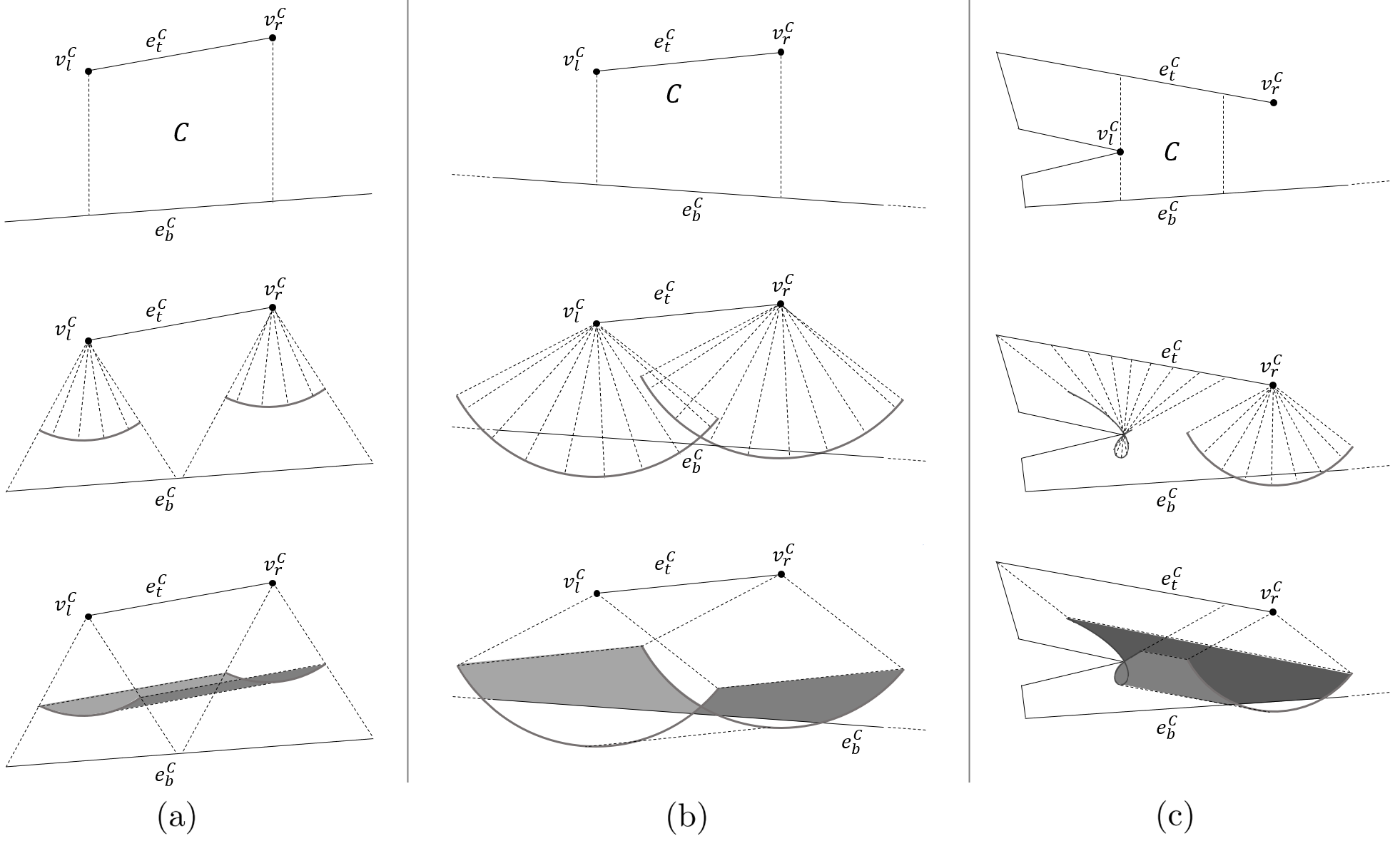}
	\caption{ \label{fig:planarShape} \sf
		Examples of the planar projection of possible configurations of a single cell with a fixed reading $d$: (a) Both limiting vertices lie on the top edge, the resulting curves are circular arcs and do not intersect the bottom edge.
		(b) The limiting vertices and arcs are similar to (a), but the result intersects the bottom edge.
		(c) One limiting vertex lies on the top edge and one does not, resulting in a circular arc and a conchoid.}
\end{figure}

We handle each of the two sub-intervals separately.
For each sub-interval (or for the entire interval $\Theta^C$ when no splitting is required) $[\theta_1,\theta_2)$, we describe all the points on the line segments $(p^{C}_{l}(\theta), p^{C}_{r}(\theta))$ for any $\theta \in [\theta_1,\theta_2)$ by a shape defined by the vertices $(v_1,v_2,v_3,v_4) = (p^{C}_{l}(\theta_1), p^{C}_{l}(\theta_2), p^{C}_{r}(\theta_2), p^{C}_{r}(\theta_1))$ and the edges $p^{C}_{l},p^{C}_{r}$ between $(v_1,v_2),(v_3,v_4)$ respectively and straight line edges between $(v_2,v_3),(v_4,v_1)$.
The union of the two shapes is the desired two-dimensional region in the workspace the sensor might be in.
The only obstacle we need to consider is the bottom edge $e_b^C$ of the trapezoid, and we can simply intersect the result with the upper half-plane defined by the line through $e_b^C$ to get the final result of the cell.
See \Cref{fig:planarShape} for an illustration.

\subsection{Suboptimality of the RTD-Based Solution}
\label{subsec:single-measurement-rtd-suboptimality}

The  RTD-based data structure is output sensitive with respect to the RTD decomposition, and in this section we explain why the data structure is not output sensitive with respect to an optimal algorithm.
In fact the gap from optimality is large: $\Theta(n)$ as shown in the following example.

\begin{figure}[h] \centering
	\includegraphics[page=4, scale=1.2]{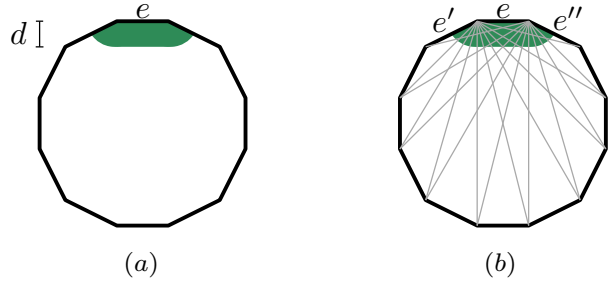}
	\caption{ \label{fig:single-suboptimality} \sf
		(a) An example of a regular (convex) polygon workspace with $n$ vertices, and a given distance measurement $d$.
		The possible configurations for a sensor's measuring ray to hit one of the edges, denoted $e$, is shown in green.
		The complexity of that area is $O(1)$ and therefore the result for the whole workspace can be described using $O(n)$ space.
		(b) The RTD cells with top edge $e$ are shown, there are $\Theta(n)$ of them.
		Each one of these cells will have a non-empty result. Therefore the complexity of the result for the whole workspace will be $\Theta(n^2)$, which is greater than the optimal by a factor $n$. }
\end{figure}

Consider \Cref{fig:single-suboptimality}, where the workspace is a regular (convex) polygon with $n$ edges, and the sensor reading is a value $d$, which is small relative to the diameter of the workspace.
The combinatorial complexity of the union of all the possible configurations yielding a reading $d$ is $\Theta(n)$, whereas the sorted-list structure will report an output of size $\Theta(n^2)$.
The reason is that all the answers with a fixed ceiling edge $e$ are fragmented by the RTD decomposition into $\Theta(n)$ cells each, while $d$ is too small to be affected by the floors of most of these cells, and an optimal output-sensitive answer to the query would have reported the answer per fixed ceiling edge in $O(1)$ fragments, each of constant size.

\section{Output-Sensitive Data Structure for Single Measurement}
\label{sec:single-measurement-output-sensitive}

We described a data structure at \Cref{sec:single-measurement-rtd}, which is output sensitive \emph{with respect to the RTD}, but as mentioned in \Cref{subsec:single-measurement-rtd-suboptimality}, it is far from answering queries optimally.
In this section we present a more involved data structure, which provides output-sensitive answers to single-measurement queries.

An edge $e$ of the workspace and a fixed distance $d$ induce a ruled surface $\Sigma(e,d)$,
which is the locus of all points $(p,\theta)\in \Re^2\times \mathbb{S}^1$ such that there is a segment $\overline{pq}$ in $\Re^2$ of length $d$, with $q\in e$, such that the vector $\overrightarrow{pq}$ makes an angle $\theta$ with the positive $x$-axis, and such that $\overline{pq}$ is contained in the half-plane defined by the line supporting $e$ and locally near $e$ facing the interior of the workspace.
We consider $e$ to be one-sided---we are only interested in the side of $e$ facing the interior of the workspace.
The union of points on $\Sigma(e,d)$ for which $\overline{pq} \in W$, defines a collection of surface patches, which we denote by $\sigma(e,d)$.
For a single measurement $d$, and for every edge $e$ on the boundary of $W$, we need to report $\sigma(e,d)$ unless $\sigma(e,d)=\emptyset$.
The combinatorial complexity of these surface patches is the total number of vertices, edges (maximal arcs of low degree algebraic curves) and faces
that define the patches.
This complexity is a lower bound on the query time of any algorithm for computing the required preimage.

We will build the data structure by varying $d$ from $0$ to $D$, where $D$ is the diameter of the workspace, and updating the result $\sigma(e,d)$ only when it changes combinatorially, for any edge $e$.
The vertices and edges of the result of $\sigma(e,d)$ will be stored in an interval tree~\cite[Chapter~10]{DBLP:books/lib/BergCKO08} $\Delta(e)$, where the interval is the set of $d$ values for which the specific result edge prevails, so it is possible to retrieve the relevant features for any query $d$ value.

\subsection{Result Vertices, Edges and Their Changes}
Fix an edge $e$.
We now embed critical surfaces in $\Re^2\times \mathbb{S}^1$ such that for a given query value $d$, they will induce a subdivision on $\Sigma(e,d)$, some of whose faces and \emph{all} its vertices and edges, constitute the desired $\sigma(e,d)$.
For each vertex in the locally-free half-plane of $e$ let $\chi_e(v)$ be the surface which is the union of the following rays: for every $\theta$, the ray emanating from $v$ in direction $\theta+\pi$.
For each edge $e'$ let $\psi(e')$ be the surface which is simply the edge $e'$ for every $\theta$, namely it is a vertical rectangle in $\Re^2\times \mathbb{S}^1$.

We call the vertices and edges of $\sigma(e,d)$ the \textit{result vertices} and \textit{result edges} (for edge $e$ and measurement $d$) to avoid ambiguity with the workspace features.
The result vertices are created by intersecting $\Sigma(e,d)$ with two critical surfaces.
By the general position assumption, at most two surfaces $\chi_e(v_i)$ intersect in a single point---in fact in a single ray.
Similarly, at most two surfaces $\psi(e_i)$ intersect at a point (in a vertical line segment).
Two surfaces $\psi(e_1),\psi(e_2)$ can intersect $\Sigma(e,d)$ at the same point, but in such case there will be another surface $\chi_e(v)$ involved where $v$ is the workspace vertex, which is the common endpoint of $e_1$ and $e_2$.
Therefore, the result vertices are obtained in two types of intersections:
\begin{itemize}
	\item
	      Type (i): Two $\chi_e(\cdot)$ surfaces intersect $\Sigma(e,d)$.
	\item
	      Type (ii): A single $\chi_e(\cdot)$ surface intersects $\Sigma(e,d)$, along with some other $\psi(\cdot)$ surface.
\end{itemize}

\begin{figure}[h] \centering
	\includegraphics[page=1, scale=0.7]{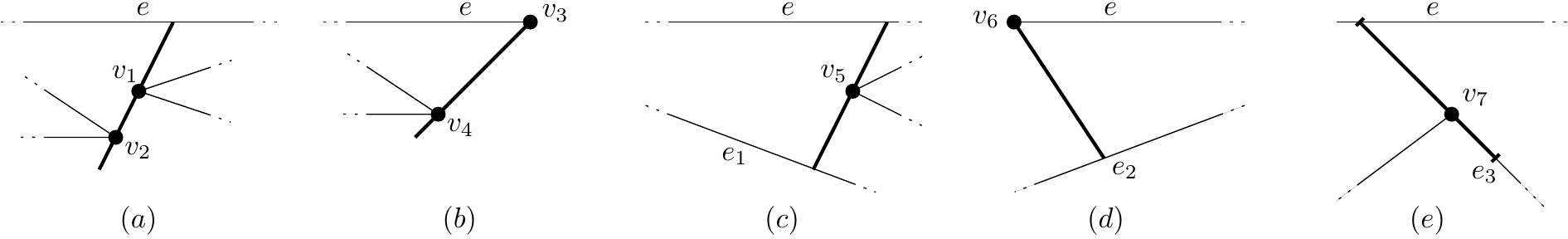}
	\caption{ \label{fig:result-vertices} \sf
		Examples of result vertices of type~(i) in (a),(b) and of type~(ii) in (c),(d),(e).
		The bold segment has length $d$, and represents a single point in the configuration space, which is a vertex in the result $\sigma(e,d)$.
		A result vertex created by surfaces of two vertices of the workspace~(a), where one of them can be an endpoint of $e$~(b).
	}
\end{figure}

\begin{figure}[h] \centering
	\includegraphics[page=1, scale=1]{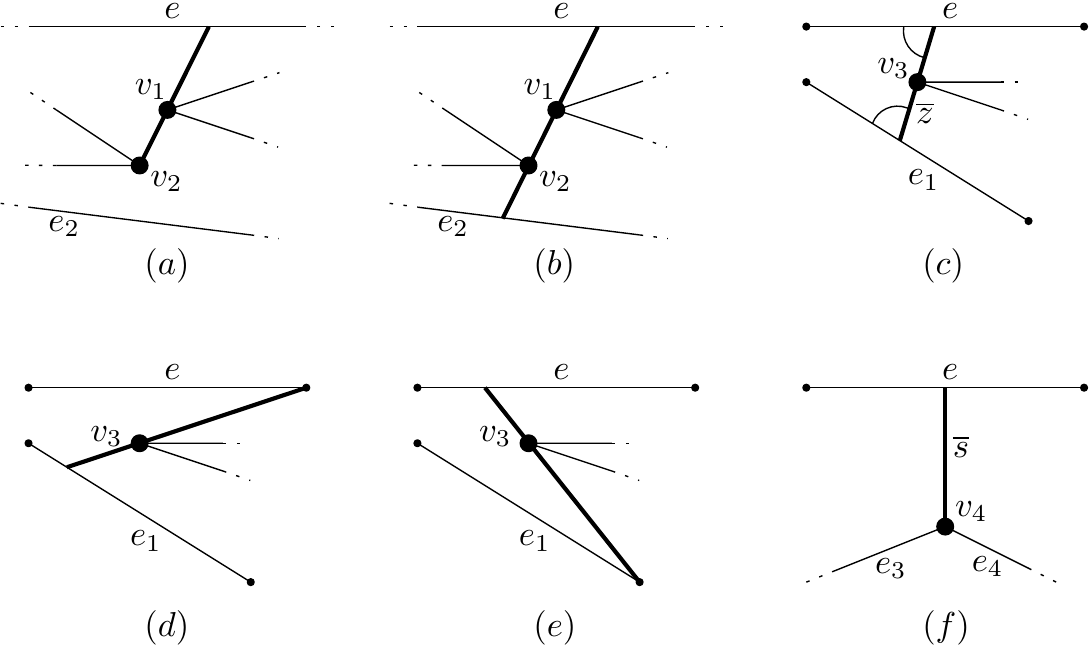}
	\caption{ \label{fig:result-vertices-comb-changes} \sf
		Examples of configurations and values of $d$ in which the result changes combinatorially.
		(a)(b) The unique minimum and maximum values of $d$ in which $\Sigma(e,d), \chi_e(v_1), \chi_e(v_2)$ create a result vertex.
		(c)(d)(e) The unique minimum and two maximal values of $d$ in which $\Sigma(e,d), \chi_e(v_3), \phi(e_1)$ create a result vertex.
		The minimum is achieved when the segment of length $d$ makes two equal angles with $e,e_1$.
		(f) The minimum value of $d$ in which $\Sigma(e,d), \chi_e(v_4), \phi(e_3)$ and $\Sigma(e,d), \chi_e(v_4), \phi(e_4)$ induce result vertices.
	}
\end{figure}

The overall scheme for preprocessing a single edge $e$ is as follows.
We will sweep the surface $\Sigma(e,d)$ over the space $\Re^2\times \mathbb{S}^1$, from $d=0$ to $d=D$, where $D$ is the diameter of the workspace $W$.
(Notice that the sweeping surface itself changes throughout the process; however, it changes continuously.)
At all times during the spatial sweep we will record the edges that appear on the boundary of $\sigma(e,d)$ in an interval tree $\Delta(e)$.
The exact geometric curve of each edge is not explicitly stored as it changes as $d$ varies, rather the workspace features that define the curve are stored, which enables us to compute it exactly for a specific $d$ in $O(1)$ time, together with the attribute on which side of an edge does the result lie.
Given a query measurement $d$, using simple auxiliary data structures to be described later, we will efficiently retrieve the $k_e$ edges of $\sigma(e,d)$, and in $O(k_e\log k_e)$ time will construct a DCEL describing the result $\sigma(e,d)$.
The events of the sweep are those where the surface patches $\sigma(e,d)$ change combinatorially.
These in turn happen at
all the minimum and maximum values of $d$ in which each result vertex exists.
There are five types of such events:
\begin{itemize}
	\item[(I)]
	      The minimum value of $d$ at which a result vertex of type (i) involving two vertices $v_1,v_2$ exists is the length of the line segment from $v_2$ to $e$ which goes through $v_1$, assuming $v_2$ is further away from $e$ than $v_1$.
	      See \Cref{fig:result-vertices-comb-changes}(a).

	\item[(II)]
	      The maximum value of $d$ in which a result vertex of type (i) involving two vertices $v_1,v_2$ exists is the length of the maximal line segment passing through $v_1,v_2$, hitting $e$ at one end and another edge of the workspace at its other end.
	      See \Cref{fig:result-vertices-comb-changes}(b).

	\item[(III)]
	      The minimum value of $d$ in which a result vertex of type (ii) involving a vertex $v_3$ and an edge $e_1$ exists.
	      If $v_3$ is the closer endpoint of $e_1$ to $e$, $d$ is the length of the segment starting at $v_3$ extending $e_1$ (along the line supporting $e_1$) and hitting $e$ (see \Cref{fig:result-vertices}(e)).
	      If $v_3$ is a reflex vertex of the workspace that is not an endpoint of $e_1$, then $d$ is the length of the minimum line segment $s$ passing through $v_3$, hitting $e$ and $e_1$ at its endpoints---this is the unique such segment that makes an equal angle with both edges, i.e., $\measuredangle{se}=\measuredangle{se_1}$ (see \Cref{fig:result-vertices-comb-changes}(c)).

	\item[(IV)]
	      The two maximal values of $d$ in which a result vertex of type (ii) involving a vertex $v_3$ and an edge $e_1$ exists are the maximal lengths of a line segment passing through $v_3$, hitting $e$ and $e_1$ at its endpoints---there are two such maximal lengths.
	      See \Cref{fig:result-vertices-comb-changes}(d,e).

	\item[(V)]
	      The distance of a vertex $v$ to $e$, when the shortest line segment from $v$ to $e$ is contained in the workspace.
	      See \Cref{fig:result-vertices-comb-changes}(f).
\end{itemize}

We wish to find all the values of $d$ for which the result $\sigma(e,d),\forall e$ is changing combinatorially when $d$ is varying from $0$ to $D$, the diameter of $W$.
To that end we carry out a global preprocessing stage, where we collect \textit{combinatorial change tuples} of the form $(e, \theta, d, U)$, where $d$ is a threshold in which the result of edge $e$ is changing combinatorially, and $U$ is the set of workspace vertices involved in the change.
Collecting all these tuples is done similarly to computing the RTD, by carrying out radial sweeps through all the vertices of the workspace, considering only pairs of visible vertices by computing the visibility graph in $O(E+n \log n)$ time (where $E$ is the number of edges in the visibility graph of $W$), and requires $O((E+n) \log n)$ time and $O(E+n)$ space.
The details of these sweeps are technical, and we defer their full description to \Cref{subsec:comb-changes-calc}, where it is also shown that at most $O(E)$ such tuples are exist and indeed collected at the preprocessing stage.

\subsection{Combinatorial Result Change Handling}
\label{subsec:Combinatorial-result-change-handling}

All the tuples with the same first parameter $e$, sorted in increasing $d$ value, are the events of the spatial sweep with the surface $\Sigma(e,d)$.
We compute the result $\sigma(e,d)$ for infinitesimal values of $d$, by considering only $e$ and its two adjacent edges.
From this point on, result edges will be inserted or removed only when handling combinatorial change tuples.
We handle each combinatorial change $(e, \theta, d, U)$ according to the event type.
The details of these handling is technical, and we focus on the handling of type~I changes in the this section.
The full description of the handling of changes is given in \Cref{app:comb-change-ops}.

\begin{figure}[h] \centering
	\includegraphics[page=1, scale=0.95]{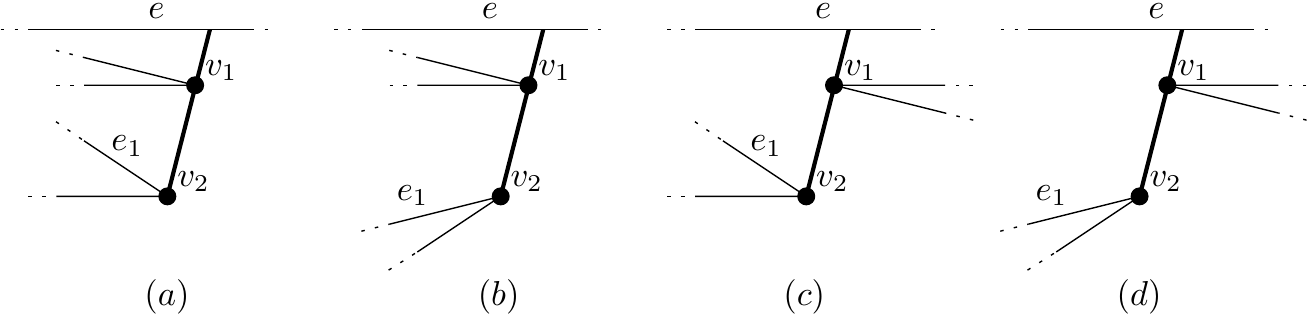}
	\caption{ \label{fig:result-vertices-comb-change-I-subcases} \sf
		Examples of the sub-cases arising in the handling of combinatorial changes of type~I.
		The bold line segments are each of length~$d$.
		There is a result edge defined by $e,e_1$ for $d'$ infinitesimally smaller than $d$ in (a),(c) and there is no such result edge in (b),(d).
		Edges of both $v_1,v_2$ are on the same halfplane relative to $\overline{v_1v_2}$ in (a),(b) and in opposing sides in (c,d).
	}
\end{figure}

\begin{description}
	\item[Combinatorial change of type~\textbf{\textnormal{I}}]:
	      $d$ is the minimum value in which a result vertex of type~(i)
	      involving two vertices $v_1,v_2$ exists (namely an intersection of two $\chi_e(\cdot)$ surfaces intersects $\Sigma(e,d)$), where $v_1$ is the vertex closer to $e$.
	      We say that a vertex is a \textit{reflex vertex relative to a line (resp.\ segment)} if both of its incident edges are in the same half-plane created by the line (resp.\ the line supporting the segment).
	      We will assume that $v_2$ is a reflex vertex relative to $\overline{v_1v_2}$; otherwise, $d$ is also the maximum value in which such a result vertex exists, and we will handle it as a change of type~II (see \Cref{fig:result-vertices-comb-changes-sweep}(a) for such a scenario).
	      $v_1$ is clearly a reflex vertex relative to $\overline{v_1v_2}$.
	      Let $e_1$ be the edge of $v_2$ visible from $v_1$.
	      We say that a result edge $\eta'$ is defined by $e$ and a vertex $v$ when $\eta'$ is the union of configurations where a rod of length $d$ is sliding on $e$ and $v$.
	      We say that a result edge $\eta''$ is defined by $e$ and another edge $e_1$ when $\eta''$ is the union of configurations where a rod of length $d$ is sliding with one endpoint on $e$ and another endpoint on $e_1$.
	      The limits $\alpha, \beta$ of the angle range $[\alpha, \beta]$ throughout which a result edge prevails, can be specified by a fixed angle each, or by the labels of the workspace features that induced these angles for a specific $d$ value.
	      The handling operations depend on two predicates:
	      whether the edges of both $v_1,v_2$ are in the same halfplane relative to $\overline{v_1v_2}$ or not, and whether a result edge defined by $e,e_1$ exists or not in $\sigma(e,d')$ for $d'$ is
	      infinitesimally smaller than $d$.
	      The sub-cases are illustrated in \Cref{fig:result-vertices-comb-change-I-subcases}.
	      We present here the operations performed in one sub-case:
	      The edges of both $v_1,v_2$ are in the left halfplane relative to $\overline{v_1v_2}$,
	      and a result edge $\eta_1$ defined by $e,e_1$ exists in the angle range $[\alpha, \gamma]$ for some $\alpha$, where $\gamma$ depends on $d$ and it is defined as the angle that the line segment passing through $v_1$ with varying length $d$ with its endpoints on $e,e_1$ makes with the $x$-axis.

	      There must be another result edge $\eta_2$ defined by $e,v_1$ in the angle range $[\gamma, \beta]$ for some $\beta$.
	      We perform the following operations:
	      \begin{itemize}
		      \item Find and remove $\eta_1$ and $\eta_2$.
		      \item Add a result edge defined by $e,v_2$ in the angle range $[\delta, \theta]$ where $\delta \leq \theta$ is the angle of the line segment with varying length $d$ with its endpoints on $v_2$ and $e$ (there are two such segments and we choose one of them by the constraint $\delta \leq \theta$).
		            Recall that $\theta$ is a parameter of the combinatorial change tuple.
		      \item Add a result edge defined by $e,v_1$ in the angle range $[\theta, \beta]$.
	      \end{itemize}
\end{description}

It is important to note that we do not maintain face objects of the subdivision $\sigma(e,d)$,and  we only maintain the edges of this subdivision.
This allows us to handle combinatorial changes of type~V by creating a new hole inside an existing face without locating the enclosing face.
For handling most combinatorial change types, it is required to find an existing edge in the list of current edges on $\sigma(e,d)$, based on the tuple information.
To that end, we maintain auxiliary balanced search trees for each pair of edge and vertex $T(e_i,v_j)$ for all $e_i$ and $v_j$, and $T(e_k,e_\ell)$ for each pair of edges $e_k,e_\ell$.
The search trees $T(e_i,v_j)$, $T(e_k,e_\ell)$ will contain all the edges currently in $\sigma(e,d)$ that are defined by $(e_i,v_j)$ and $(e_k,e_\ell)$, respectively.
The key for the auxiliary trees is the smallest angle in the angle interval in which the edge exists.
The angles are not explicitly stored in the tree (since they change as $d$ varies), rather only the order of the edges is maintained.
This is possible because multiple edges defined by the same features exist in disjoint ranges of angles, and their order does not change.
When handling a combinatorial change, the result edges can be found using simple $O(\log n)$-time search with the angle $\theta$ in the relevant trees, and newly created edges are inserted into these auxiliary data structures, in $O(\log n)$ time total.

\subsection{Storing the Result Edges for Efficient Queries}

For each edge~$e$, the interval tree $\Delta(e)$ is constructed independently using the associated combinatorial changes.
Let $\nu_e$ be the number of combinatorial changes involving the edge~$e$.
Construction of a single data structure $\Delta(e)$ requires $O(\nu_e \log n)$.
The overall number of events over all edges is $O(E)$ and therefore it takes total $O(E \log n)$ time to construct the data structures for all the edges altogether.
Finally, we also compute for every edge $e$ the maximum $d$ for which $\sigma(e,d) \neq \emptyset$; we denote this maximum by $m(e)$.
We will then keep these interval trees maxima sorted in decreasing order in a list $L$.
The preprocessing requires $O((E+n) \log n)$ time and $O(E)$ space in total.

During a query for a single measurement $d$, we iterate over the data structures in $L$ by their sorted order as long as $m(e) \geq d$, ensuring each data structure we examine will have a non-empty result, excluding possibly the last one.
(Notice that if a structure has non-empty result for $d'$ then it also has a non-empty result for any $d<d'$.)
For each data structure $\Delta(e_i)$ that has a non-empty result, we query the interval tree for the result edges, in time $O(\log n +k_i)$, where $k_i$ is the number of result edges found.
For each of result edge, we compute its exact arc, which depends on the query specific $d$ value.
Then using a standard algorithm,\footnote{See~\cite[Section~2.3]{DBLP:books/lib/BergCKO08} for details; it is described there for completing the overlay of two subdivisions, but applies equally well here after a preprocessing step where we form boundary cycles of the DCEL.} we construct the complete desired DCEL in $O(k \log k)$ time, where $k=\Sigma_{i=1}^n k_i$ is the number of result edges in total.
The result faces can be identified since, as noted above, each edge stores which of its sides (halfedges in the DCEL) is contained in the result.
The total running time of the query is $O(k \log n)$.

\begin{theorem}
	Given a polygonal workspace with a total of $n$ vertices, a data structure can be constructed in time $O((E+n) \log n)$ and $O(E+n)$ space, where $E$ is the number of edges in the visibility graph of $W$, such that, given a query real parameter $d>0$, we can report all possible configurations yielding depth measurement $d$ in time $O(k \log n)$, where $k$ is the number of vertices and maximal arcs of low algebraic degree appearing on the boundary of the surface patches that constitute the result.
\end{theorem}

\subsection{Computing the Combinatorial Change Tuples}
\label{subsec:comb-changes-calc}

We wish to find all the values of $d$ for which the result $\sigma(e,d),\forall e$ is changing combinatorially when $d$ is varying from $0$ to $D$, the diameter of $W$.
To that end we carry out a global preprocessing stage, where we collect \textit{combinatorial change tuples} of the form $(e, \theta, d, U)$, where $d$ is a threshold in which the result of edge $e$ is changing combinatorially, and $U$ is the set of workspace vertices involved in the change.

Collecting all these tuples is done similarly to computing the RTD.
First, the visibility graph of $W$ is computed in $O(E+n \log n)$ time~\cite{DBLP:journals/siamcomp/GhoshM91}, where $E$ is the number of edges in the visibility graph of $W$.
Let ${\cal E}$ denote the set of edges in the visibility graph of $W$.
Then, we sort the set of ordered pairs $ \bigcup_{\{u,v\} \in {\cal E}} {\{(u,v),(v,u)\}}$ by the slope of the line oriented at $v_1$ directed at $v_2$ of each ordered pair $(v_1,v_2)$, and perform from each vertex $v$ of the workspace a rotational plane sweep, in two antipodal directions simultaneously.
The other visible vertices of the workspace are the events of the sweep, and the first edge intersecting each imaginary sweep ray is the state of the sweep.
(There is no need to store all edges that intersect the ray, as we already know the visibility graph).
Consider an event at a vertex $u \neq v$.
The event is handled differently depending on whether $u$ and $v$ are \textit{reflex} vertices relative to the line $\overline{vu}$.
Recall that we say that a vertex is a \textit{reflex vertex relative to a line} if both of its edges are in the same halfplane bounded by the line.

\begin{figure}[h] \centering
	\includegraphics[page=1, scale=0.95]{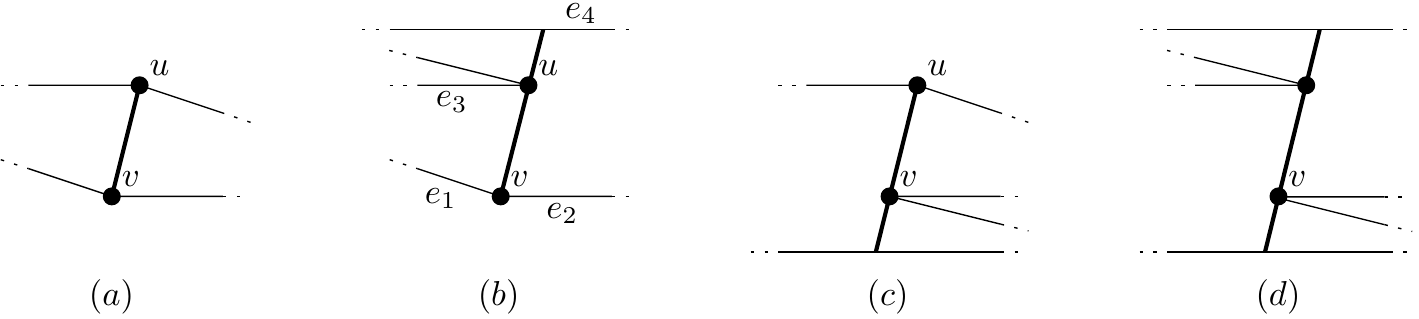}
	\caption{ \label{fig:result-vertices-comb-changes-sweep} \sf
		Examples of scenarios arising when handling an event of vertex $u$ during a rotational plane sweep around a vertex $v$.
		(a) Both $u$ and $v$ are not reflex vertices relative to $\overline{uv}$.
		(b)(c) Only one of $u,v$ is a reflex vertex relative to $\overline{uv}$.
		(d) Both $u$ and $v$ are reflex vertices relative to $\overline{uv}$.
	}
\end{figure}

We will cover here in details the case where $u$ is a reflex vertex relative to $\overline{vu}$ and $v$ is not; the other cases are handled in an analogous fashion:
Let $e_1, e_2$ be the edges to the right and left of $v$, respectively, relative to the direction $\overrightarrow{uv}$.
Let $e_3$ be the edge of $u$ that is visible from $v$ and let $e_4$ be the edge that the ray originating at $v$ in the direction $\overrightarrow{vu}$ hits after passing through $u$ (see \Cref{fig:result-vertices-comb-changes-sweep}(b)).
Let $\theta$ be the angle of the direction $\overrightarrow{vu}$, relative to the $x$-axis,
let $dist(v, u)$ be the distance between $v$ and $u$,
and let $dist(v,u,e_4)$ be the length of the segment originating at $v$, passing through $u$ and hitting $e_4$.
Add the following combinatorial change tuples:
\[ \begin{matrix}
		(e_1, \theta+\pi, dist(v, u), \{v, u\}), &
		(e_1, \theta+\pi, dist(u, v, e_4), \{v, u\}), \\[0.5em]
		(e_2, \theta+\pi, dist(v, u), \{v, u\}), &
		(e_2, \theta+\pi, dist(u, v, e_4), \{v, u\}), \\[0.5em]
		(e_3, \theta, dist(v, u), \{v, u\}),     &
		(e_4, \theta, dist(u, v, e_4), \{v, u\}).
	\end{matrix} \]
The above six combinatorial change tuples are the changes of types that include two vertices, namely changes of types I, II, and IV.
To find the changes including only a single vertex, during the rotational sweep around the vertex $v$, we consider the vertices $u_1, u_2$ of two consecutive events.
Let $\theta_1,\theta_2$ be the angles the directed lines $\overrightarrow{vu_1}, \overrightarrow{vu_2}$ make with the positive $x$-axis, respectively.
Denote by $e_1,e_2$ the edges that are hit by the two antipodal rays of the sweep for any angle $\theta \in (\theta_1,\theta_2)$ and $\theta + \pi$ respectively.
Denote by $\overline{z}$ the line segment that passes through $v$ and makes equal angles when hitting $e_1,e_2$. The segment $\overline{z}$ is directed from its endpoint on $e_1$ to its endpoint on $e_2$.
If the angle $\measuredangle{\overline{z}}$ that the segment $\overline{z}$ makes with the positive $x$-axis, is in the range $(\theta_1,\theta_2)$, add the combinatorial tuple change of type III $(e_1, \measuredangle{\overline{z}}, ||\overline{z}||, \{v\})$, where $||\overline{z}||$ is the length of $\overline{z}$ (see \Cref{fig:result-vertices-comb-changes}(c) for illustration).
Denote by $\bar{s}$ the shortest line segment between $v$ and $e_1$.
If the angle $\measuredangle{\bar{s}}$ that $\bar{s}$ makes with the $x$-axis is  in the range $(\theta_1,\theta_2)$, add the combinatorial tuple change of type V $(e_1, \measuredangle{\bar{s}}, ||\bar{s}||, \{v\})$ (see \Cref{fig:result-vertices-comb-changes}(f) for illustration).

We perform such a rotational sweep around each vertex of the workspace and add all the combinatorial change tuples.
Each time an event is handled, a constant number of combinatorial change tuples are added, therefore the total number of such combinatorial change tuples is $O(E)$.
The time complexity of this preprocessing stage is $O((E+n) \log n)$.

\section{Localization by Antipodal Measurements}
\label{sec:antipodal-measurements}

An important variant of the problem is a sensor that takes two different distance measurements from the same location, in opposite directions.

We obtain one measurement $d_1$ in (unknown) direction $\theta_1$ and a second measurement $d_2$ in direction $\theta_2=\theta_1+\pi$.
We seek to efficiently compute the intersection of the C-space preimages that correspond to each measurement.
A nice, mitigating property of this setting is that the two distance measurements are necessarily taken to two \emph{distinct} edges of the workspace.

Two data structures are presented in this section:
(i) a simpler, RTD-based data structure, which is output sensitive with respect to the RTD, and
(ii) an output sensitive data structure, which uses the RTD cells as intermediate building blocks, but then modifies the decomposition cells by unions and splits of the cells into a more economical collection.

\subsection{RTD-Based Data Structure for Antipodal Measurements}
\label{subsec:antipodal-measurements-rtd}

In this section, a solution based on the decomposition described in \Cref{sec:RTD} is used, where the orientation $\theta$ is the orientation of the first measurement.
The data structure achieves a query time of $O(\log n + k)$, where $k$ is the number of RTD cells that contains a result configuration.

For a fixed orientation $\theta$, the cross-section of an RTD cell $C$ contains one potential sensor pose (assuming the top and bottom edges of the cell are not parallel), and we can calculate it by finding the segment of length $d_1+d_2$ with its endpoints on the cell's top and bottom edges. Which in turn we do by solving $O^{C}(\theta,x)=d_1+d_2$, and extracting the point along the segment obtaining this opening that is at distance $d_1$ from the top edge (and hence at distance $d_2$ from the bottom edge).
For each such cell $C$ we compute the locus $\Gamma(C)$ of potential sensor poses, which is a curve parameterized by the orientation of the first measurement.

In order to store the cells of the decomposition so as to be able to efficiently retrieve the cells $C$ with non-empty loci $\Gamma(C)$, we now need to also compute the minimum opening of a cell $\ominc{C}$, which is defined analogously to $\omaxc{C}$ (cf.~Section~\ref{sec:single-measurement-rtd}).
Given a pair of antipodal measurements $(d_1,d_2)$, a cell $C$ has a non-empty set of loci if and only if $\ominc{C} \leq d_1+d_2 \leq \omaxc{C}$.
Each cell is associated with an interval $[\ominc{C},\omaxc{C}]$, and we construct an interval tree over these $O(E)$ intervals.
Given the measurements
$(d_1,d_2)$, we extract all the intervals that contain the value $d_1+d_2$, and their corresponding cells contribute potential poses to the final answer.

The loci $\Gamma(C)$ constitute an arc of an ellipse with its center at the (possibly imaginary) intersection of the lines supporting the top and bottom edges of the cell; see \Cref{app:antipodal-projection} for details.
After prepossessing in $O((E+n) \log n)$ time and using $O(E+n)$ space, a query can be answered in $O(\log n+k)$ time, where $k$ is the number of cells $C$ with non-empty $\Gamma(C)$.

\begin{theorem}
	Given a polygonal workspace $W$ with a total of $n$ vertices, an interval tree containing the RTD cells of the configuration space can be constructed in time $O((E+n) \log n)$ and space $O(E)$, where $E$ is the number of edges in the visibility graph of $W$, which is bounded by $O(n^2)$.
	Then, given two real numbers $d_1,d_2>0$, we can report all the  RTD cells that contain a possible configuration yielding antipodal depth measurement $d_1,d_2$ in time $O(\log n + k)$ where $k$ is the number of the relevant cells.
\end{theorem}

\subsection{Suboptimality of the RTD-Based Solution}
\label{sec:rtd-suboptimality}

The data structure presented at \Cref{subsec:antipodal-measurements-rtd} is output sensitive with respect to the RTD decomposition, and in this section we demonstrate why it is not output sensitive with respect to an optimal algorithm.
In fact the gap from optimality is large: $\Theta(n)$ as shown in the following example.

\begin{figure}[h] \centering
	\includegraphics[page=5, scale=1.2]{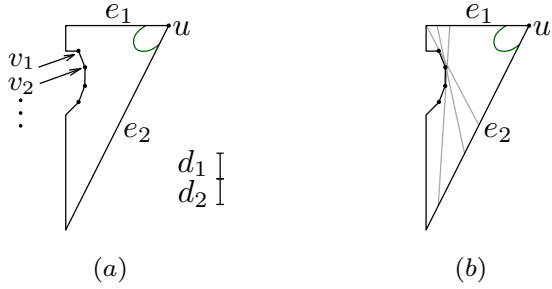}
	\caption{ \label{fig:antipodal-suboptimality} \sf
		A triangle-like workspace, with additional $\Theta(n)$ reflex vertices on the left edge.
		Given a query of small $d_1=d_2$, the possible positions a sensor can measure $e_1,e_2$ at distances $d_1,d_2$ respectively are drawn as a green arc at the top-right corner of the workspace.
		RTD  will contain $\Theta(n)$ cells with $e_1,e_2$ as top and bottom edges, and $u$ as the right limiting vertex, each with a different left limiting vertex $v_i$, shown in (b).
		Each of these cells will have a non-empty result, while the output of these two edges has $O(1)$ descriptive complexity. }
\end{figure}

Consider \Cref{fig:antipodal-suboptimality}, where the workspace is a triangle-like workspace with additional $Theta(n)$ reflex vertices on the left edge.
The angle interval in which a sensor can measure the edges $e_1$ and $e_2$ with two antipodal measurement is fragmented by the RTD decomposition into $\Theta(n)$ cells of the form $(e_1,e_2,v_i,u, \cdot)$ where $v_i$ is a different reflex vertex as a limiting left vertex of the cell.
Although this fragmentation may be needed for some values of $d_1,d_2$, for smaller such values (as shown in the figure) the change of the left limiting vertex does not affect the result.
There will be $\Theta(n)$ RTD cells with $e_1,e_2$ as top and bottom edges that will contribute to the result, while the output can be described using only $O(1)$ space.

\subsection{Output-Sensitive Data Structure for Antipodal Measurements}
\label{subsec:antipodal-output-sensitive}

The result of a query of two measurements $(d_1,d_2)$ can be naturally and economically described
by a set of angle intervals $I_{e_1,e_2}=\{[\alpha_1, \beta_1], [\alpha_2, \beta_2], \ldots\}$ for any relevant pair of workspace edges $e_1,e_2$, where for each interval the sensor reading will hit $e_1$ at any angle $\theta \in [\alpha_i, \beta_i]$ and $e_2$ at angle $\theta +\pi$, and there is only one such configuration (assuming, as we do, that there are no parallel edges in the workspace).
When considering this representation, the minimum complexity required is achieved when the intervals are maximal and disjoint.

\begin{figure}[h] \centering
	\includegraphics[scale=1]{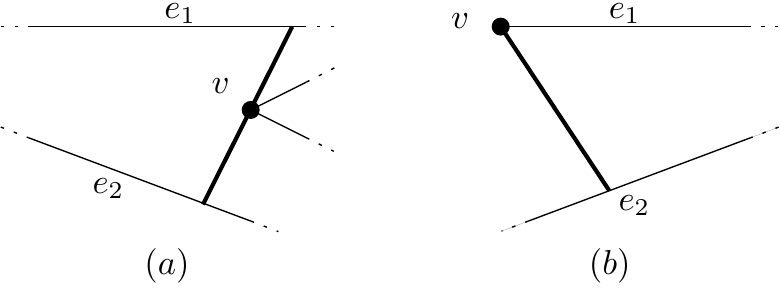}
	\caption{\label{fig:m2-res-vertices} \sf
		Examples of configurations which are vertices in the minimum-complexity result for two antipodal measurements, where the bold line segment has length $d_1+d_2$.
	}
\end{figure}

For a fixed maximal interval $[\alpha_i, \beta_i] \in I_{e_1,e_2}$, we consider the line segment of length $d=d_1+d_2$ at angle $\alpha_i$ with its endpoints lying on $e_1$ and $e_2$.
This line segment must also intersect a vertex of the workspace, either an endpoint vertex of $e_1$ or $e_2$, or some other reflex vertex of the workspace, as otherwise the angle interval would not be maximal.
The other direction also holds:
The existence of a line segment of length $d$ passing through a vertex with its endpoints lying on two different edges $e_3,e_4$ at angle $\theta$, must imply that the angle $\theta$ is included in one of the result angle intervals of $e_3,e_4$, and it must be an extreme angle in that interval.
See \Cref{fig:m2-res-vertices} for an illustration.
Therefore the minimum complexity required to describe a query result for antipodal measurements $(d_1,d_2)$ is the number of line segments of length $d=d_1+d_2$ that can be placed in the interior of the workspace such that both their endpoints lie on edges of the workspace and additionally they intersect a vertex of the workspace.

To construct a data structure that will provide such answers, we will use the RTD of the C-space in a preprocessing stage, after which we will merge results from different cells of the RTD, wherever possible.
For each RTD cell $C$, we define the minimum left opening $O^{Cl}_{min}$ as the minimum opening obtained over all the angles that $C$ exists in with the $x$-coordinate of the top left vertex of $C$.
Similarly, we define minimum/maximum left/right openings:
\[ \begin{matrix}
		O^{Cl}_{min} = \min_{\theta \in \Theta^C} {O(\theta, x^C_{lt}(\theta))} \;, &
		O^{Cl}_{max} = \max_{\theta \in \Theta^C} {O(\theta, x^C_{lt}(\theta))} \;,
		\\[1em]
		O^{Cr}_{min} = \min_{\theta \in \Theta^C} {O(\theta, x^C_{rt}(\theta))} \;, &
		O^{Cr}_{max} = \max_{\theta \in \Theta^C} {O(\theta, x^C_{rt}(\theta))} \;.
	\end{matrix} \]

Based on the relation between $d$ and the minimum/maximum left/right openings, we classify each RTD cell $C$ as one of three classes: \textit{Empty} class $Em$, \textit{Full} class $F$ or \textit{Partial} class $P$:
\[ \begin{matrix}
		Em=\{ C \mid
		d < \min\{ O^{Cl}_{min}, O^{Cr}_{min} \} \;\lor\;
		d > \max\{ O^{Cl}_{max}, O^{Cr}_{max} \} \}
		\;,    \\[1em]

		F=\{ C \mid
		O^{Cr}_{max} < d < O^{Cl}_{min} \;\lor\;
		O^{Cl}_{max} < d < O^{Cr}_{min}
		\} \;, \\[1em]

		P=\{ C \mid
		O^{Cl}_{min} < d < O^{Cl}_{max} \;\lor\;
		O^{Cr}_{min} < d < O^{Cr}_{max}
		\} \;.
	\end{matrix} \]

The members of the class $Em$ are \emph{empty} in the sense that each RTD cell in this class does not contain any valid result, as the opening is either greater or smaller than $d$ for any angle in the cell's angle interval.
The members of the class $F$ are \emph{full} in the sense that each RTD cell $C$ in this class contributes the full angle interval $\Theta^C$ to the result, since the right (resp.\ left) opening is always greater than $d$ and the left (resp.\ right) opening is always smaller than $d$, and due to the fact the opening function is linear for a fixed angle (see \Cref{subsec:rtd-opening-func}) this implies that there is some point between the left and right vertices where the opening is equal to $d$, for any angle in $\Theta^C$.
The members of the class $P$ are \emph{partial} in the sense that each RTD cell $C$ in that class contributes a part of the angle interval $\Theta^C$ to the result, since there is some angle~$\theta$ at which the right or left opening exactly equals $d$, which means that at angle $\theta+\epsilon$ (or $\theta-\epsilon$) for some small $\epsilon$ the opening for any point in the cell is not equal to $d$.

Another aspect to consider is that each cell $C$ in the partial class contains a configuration which will be a vertex in the result, as a segment of length $d$ can be placed such that its endpoints lie on $e^C_t$ and $e^C_b$ exactly on the vertices $v^C_{lt}, v^C_{rt}$, and one of these vertices must be a workspace vertex.
For each pair of edges, the angle intervals contributed by the cells of full and partial classes are pairwise interior disjoint, but may overlap at their endpoints.
By merging the angles intervals of each pair of edges to maximal intervals, the minimum-complexity output is achieved.

Let $\D$ denote the set of all minimum/maximum left/right opening of all the RTD cells, along with $\infty$ and $0$, sorted in increasing order:
\[ \D = [d_0, d_1,\ldots \mid d_i < d_{i+1}, d_i \in \{0, \infty\} \cup \bigcup_{C} {\{O^{Cl}_{min}, O^{Cl}_{max}, O^{Cr}_{min}, O^{Cr}_{max}\}}] \;. \]
For two values $d',d''$ the RTD cells classification, and therefore also the result angle intervals, are combinatorially equivalent if there exists $i$ such that $d',d'' \in [d_i, d_{i+1})$.
We can perform the classification of the RTD cells and the union of the angle intervals for each range $[d_i, d_{i+1})$ during preprocessing,
which will result in angle intervals with symbolic endpoint values, which in turn can be made concrete for any fixed value of $d$.
During query with $d_1,d_2$, we select the result corresponding to the range $[d_i, d_{i+1})$ such that $d_1+d_2 \in [d_i, d_{i+1})$, which contains maximal angle intervals and therefore optimal in size, and calculate their exact values for the specific value $d=d_1+d_2$.

Producing and storing the full result for each range $[d_i, d_{i+1})$ is expensive.
For efficiency we will again use an interval tree to store the result maximal angle intervals.
The values stored in the interval tree are the result angle intervals, and the keys of these values are the ranges of $d$ for which the values (the angle intervals) are present in the result.
Although the values are intervals themselves, they have nothing to do with the interval keys of the tree, which are, again, ranges of $d$.

Such an interval tree can be constructed in the following way.
We vary $d$ from $\infty$ to $0$ by iterating over the ranges $[d_i, d_{i+1}]$.
The result of the current value of $d$, which is a set of maximal angle intervals, is maintained at all times as a balanced search tree $R$.
Each result angle interval is stored in $R$ using a key $(e_1,e_2,\theta_1)$, where $e_1,e_2$ are the two edges the sensor could have measured, and $\theta_1$ is the minimal angle in the result interval.
This representation of the result set allows us to efficiently find the result angle intervals that are affected combinatorially when we change $d$.
We iterate over the ranges of $d$ in order, computing the result intervals of $[d_i, d_{i+1}]$ based on the result intervals of $[d_{i+1}, d_{i+2}]$.
We start with the range $[d_{|D|-1}, d_{|D|}]=[d_{|D|-1}, \infty]$, in which all RTD cells are classified as \textit{empty} and the result contains no intervals, $R=\emptyset$.
Assuming we already classified the RTD cells and calculated the result intervals for the range $[d_{i+1}, d_{i+2}]$, we assume that a single cell $C$ changes its class when considering the next range $[d_i, d_{i+1}]$.
If multiple cells change their classes, we repeat the following process multiple times.
\begin{itemize}
	\item If $C$ was full or partial relative to the previous range $[d_{i+1}, d_{i+2}]$, we remove the angle interval it contributes from $R$. We may have to split a larger interval that was created from multiple intervals to remove it.
	\item If $C$ is full or partial relative to the current range $[d_{i+1}, d_i]$, we add the angle interval it contributes to $R$. We may have to merge it with at most two additional intervals that intersect it at its endpoints.
\end{itemize}

In both cases, a constant number of insertions, deletions, splits and merges are performed.
For each interval, we store the values of $d$ in which it was added or removed from $R$, or $0$ if it was not removed after iterating over $[d_0,d_1] = [0, d_1]$.
We insert all such angle intervals into an interval tree keyed by the range of values of $d$ for which each angle interval was in $R$.

In total, the number of operations performed to build the data structure is linear in the number of RTD cells, which is $O(E)$.
Each operation on $R$ requires $O(\log n)$ time, and constructing an interval tree with $O(E)$ elements requires $O(E \log n)$ time and $O(E)$ space.
During query, the result intervals are retrieved from the interval tree in time $O(\log n + k)$, where $k$ is the number of maximal angle intervals in the result, and the exact algebraic description can be computed for each one of them in $O(1)$ time.

\begin{theorem}
	Given a polygonal workspace $W$ with a total of $n$ vertices, a data structure can be constructed in time $O((E+n) \log n)$ and $O(E)$ space, where $E$ is the number of edges in the visibility graph of $W$ (which is bounded by $O(n^2)$), such that, given a query of two measurements $(d_1,d_2)$, we can report all possible configurations yielding antipodal depth measurement $d_1,d_2$ in time $O(\log n + k)$ where $k$ is the number of maximal elliptical arcs in the result.
\end{theorem}

\section{Implementation and Experiments}
\label{sec:implementation}

We implemented the simpler data structures, which are built upon the rotational trapezoidal decomposition, both for single distance-measurement queries and for two-antipodal measurements queries (see \Cref{sec:single-measurement-rtd} and \Cref{subsec:antipodal-measurements-rtd}).
We describe our implementation details, challenges and report on some experiments that we conducted.

\begin{figure}[h] \centering
	\includegraphics[scale=0.15]{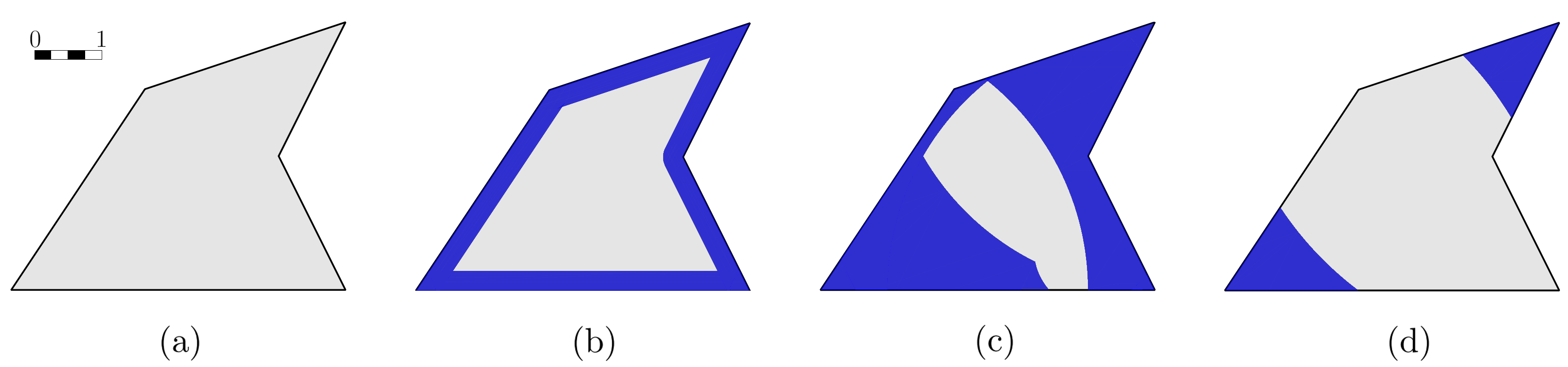}
	\caption{\label{fig:expr-m1-result} \sf
		The possible positions of a sensor that performs a single measurement in a simple polygon workspace (a) are shown in blue in (b) for $d=0.3$, in (c) for $d=4$, and in (d) for $d=5$. }
\end{figure}

\begin{figure}[h] \centering
	\includegraphics[scale=0.13]{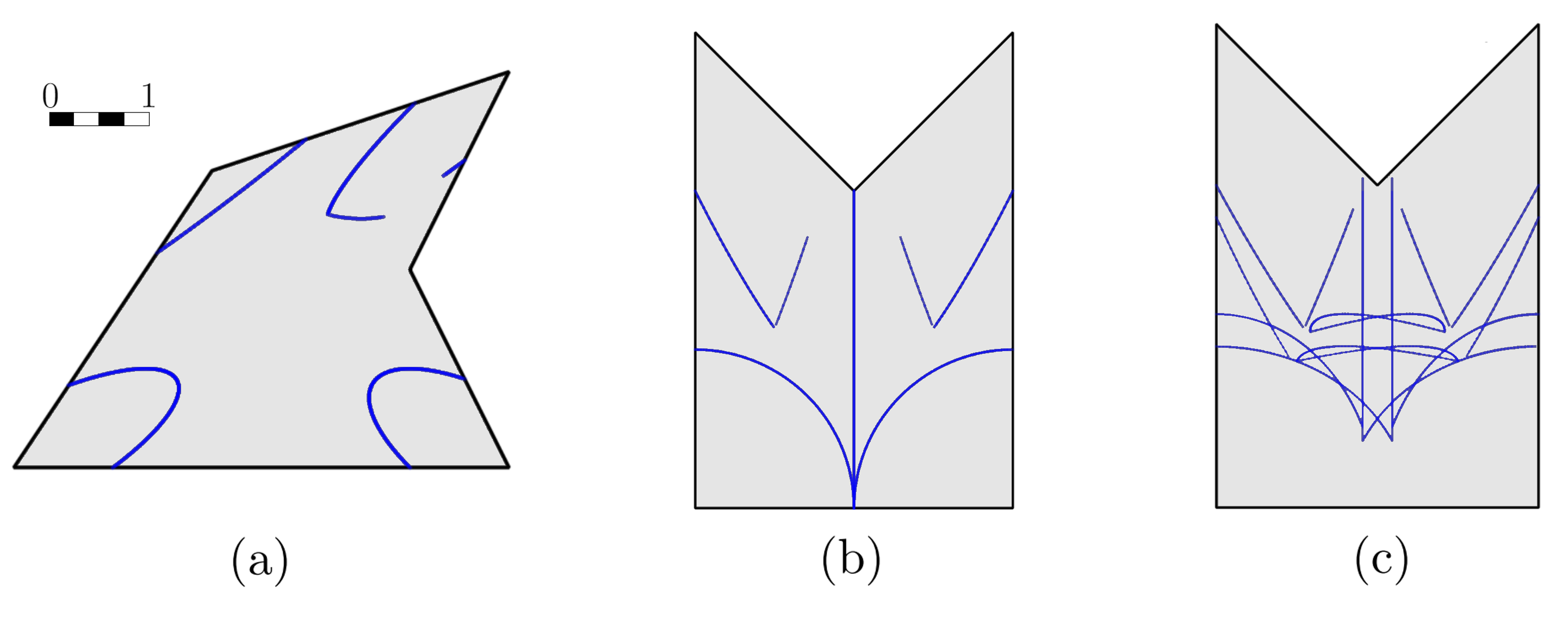}
	\caption{\label{fig:expr-m2-result} \sf
		The possible positions of a sensor that performs two antipodal measurements in simple polygons are shown in blue in (a) for $d_1=d_2=1$, (b) for $d_1=d_2=1$ and (c) for $d_1=1, d_2=1.2$.}
\end{figure}

\begin{figure}[h] \centering
	\includegraphics[scale=0.5]{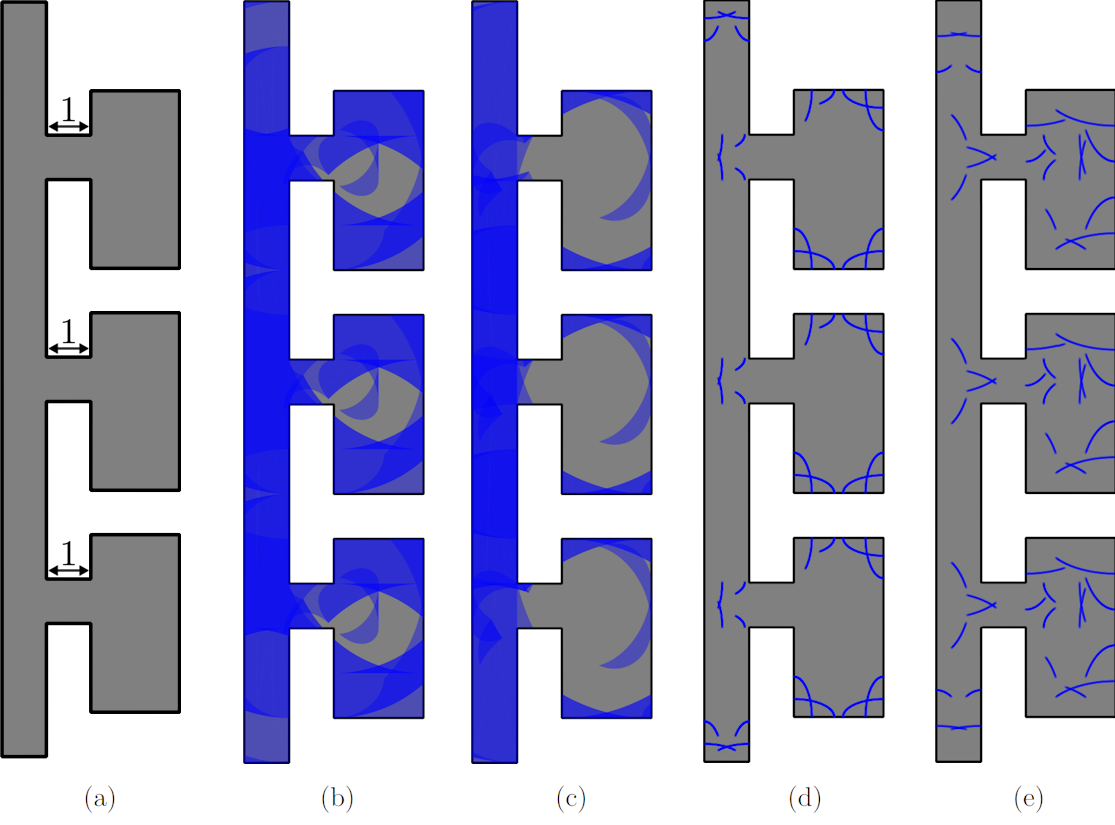}
	\caption{\label{fig:expr-symmetric} \sf The workspace at (a) is a symmetric workspace; it comprises a corridor and  three rooms, where the rooms are indistinguishable from one another for a sensor positioned in any one of them.
		In (b) and (c) the possible positions of the sensor are drawn, after it measured a single value of $d=3$ and $d=4$ respectively.
		In (d) and (e) the possible positions of the sensor are drawn, after it measured two antipodal measurements of $d_1=0.4,d_2=0.9$ and $d_1=0.8,d_2=1.6$ respectively.
		It is easy to see that if the robot is indeed located in one of the identical rooms it can not determine its position uniquely, no matter how many measurements it will perform.
		Notice that there is a range of intensity of blue color, where greater intensity corresponds to more RTD cells contributing a result to the region in question. }
\end{figure}

We implemented the structures in C++ using the CGAL library~\cite{cgal:eb-23a} and in particular the arrangements package~\cite{DBLP:books/daglib/0028679}.
The implementation is open source and available at \url{https://github.com/barakugav/FDML}.
Given a workspace, a preprocessing is performed once, and multiple queries of both types can be answered with output-sensitive complexity with respect to the RTD.
The implementation assumes general position, namely that the given workspace does not contain three vertices on the same line, three edges that intersect in a single point, etc.
In addition to the C++ code, we also implemented Python bindings on top of CGAL's bindings~\cite{DBLP:journals/corr/abs-2202-13889}, allowing for an easier use of our data structures.
Results of queries for a single measurement and for two antipodal measurements are shown in \Cref{fig:expr-m1-result} and \Cref{fig:expr-m2-result}.
Notice that  some workspace contains identical subregions, in which case a sensor located in one of these subregions can not distinguish between them, even if we allow for an unlimited number of measurements.
See \Cref{fig:expr-symmetric} for an example of such an environment.


\begin{figure}[h] \centering
	\includegraphics[scale=0.52]{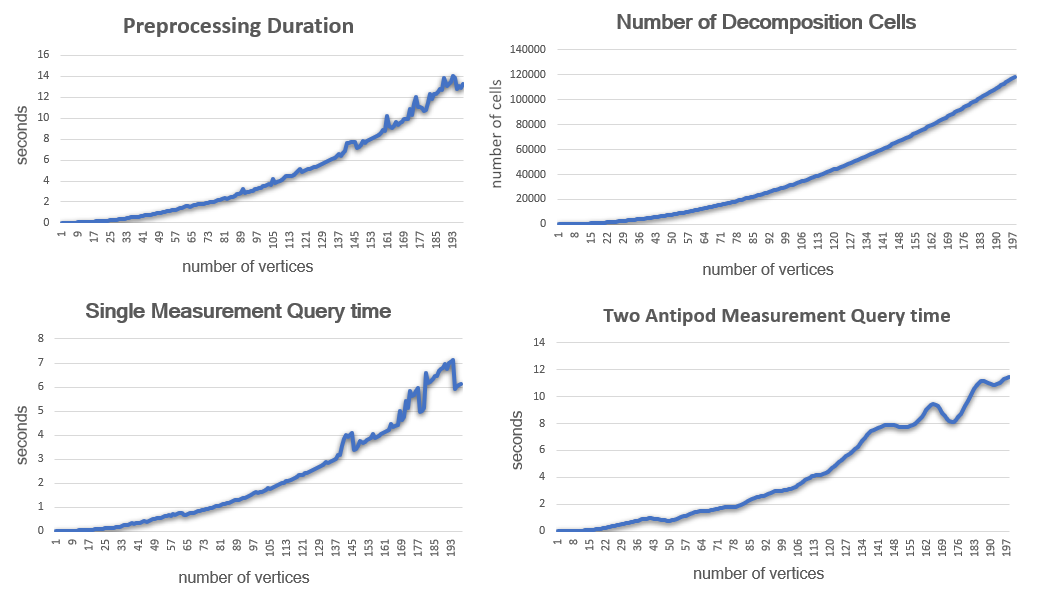}
	\caption{\label{fig:expr-perf} \sf From left to right, top to bottom. (1) The preprocessing time for a convex regular polygon with a radius of $1.0$ and increasing number of vertices.
		(2) The number of three-dimensional RTD cells that were computed.
		The query time in these polygons is shown, both for a single measurement query of $d=0.1$ at (3) and two antipodal measurements query of $d_1=d_2=0.5$ at (4).
		The preprocessing time, the number of cells and the running time for both types of queries are behaving asymptotically as $(\textit{number of vertices})^2$, as expected in a perfect polygon.}
\end{figure}

Among the  challenges we faced during the development of the software, we had to use exact number types to represent angles and points in the workspace.
Although CGAL supports these number types using the GMP or MPFR libraries, their usage has a tremendous impact on performance.
By using inexact number types such as $64$-bit floating point, the running time presented in \Cref{fig:expr-perf} can be improved by an order of magnitude.
Another challenge was to represent the result curves (projected onto the plane), as these curves are of high degree (see \Cref{app:single-projection} and \Cref{app:antipodal-projection}).
We approximate these curved using polylines, with parameterized precision.

\begin{figure}[h] \centering
	\includegraphics[scale=0.20]{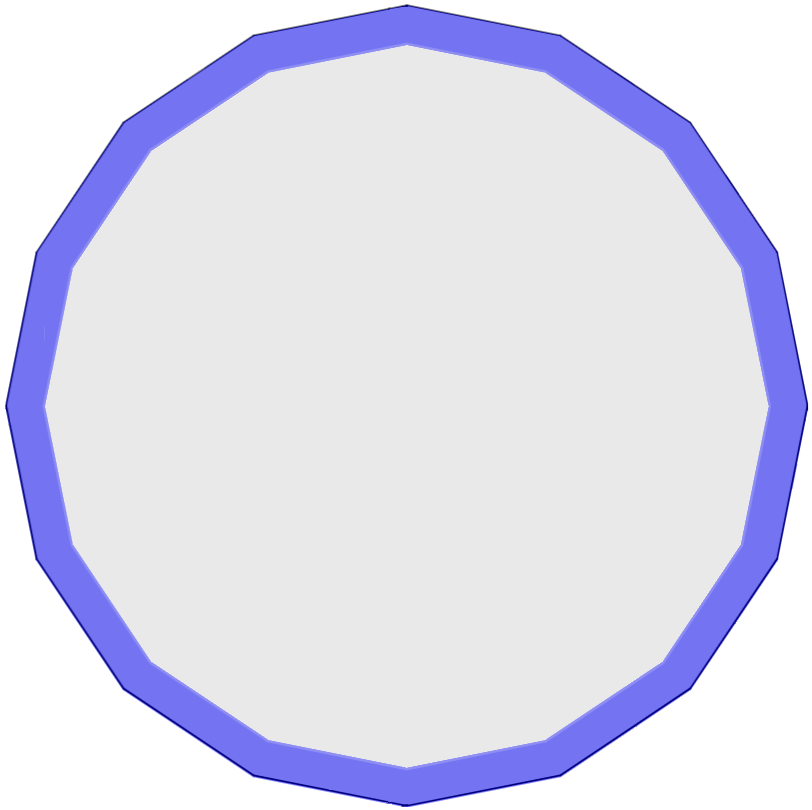}
	\caption{\label{fig:expr-regular-polygon} \sf An example of a regular convex polygon with $16$ vertices and a radius of $1$. The result of a query of a single distance measurement of $0.1$ is a 'margin' area near the polygon boundary. The RTD decomposition decomposes the C-space for the regular polygon into $O(n^2)$ cells, each containing part of the result. In our performance experiments we used such polygons with increasing number of vertices, therefore the preprocessing and query time where proportional to the squared number of vertices. }
\end{figure}

We conducted a few additional performance experiments, measuring the preprocessing time and query time for a convex regular polygon with different number of vertices; see \Cref{fig:expr-regular-polygon}.
The experiments where run on a Windows 10 64-bit machine with 11th Gen Intel Core i7-1185G7 at 3.00GHz and 16GB memory.
The results are shown at \Cref{fig:expr-perf}.

\section{Discussion and Future Work}\label{sec:conclusion}
\begin{figure}[h] \centering
	\includegraphics[scale=0.22]{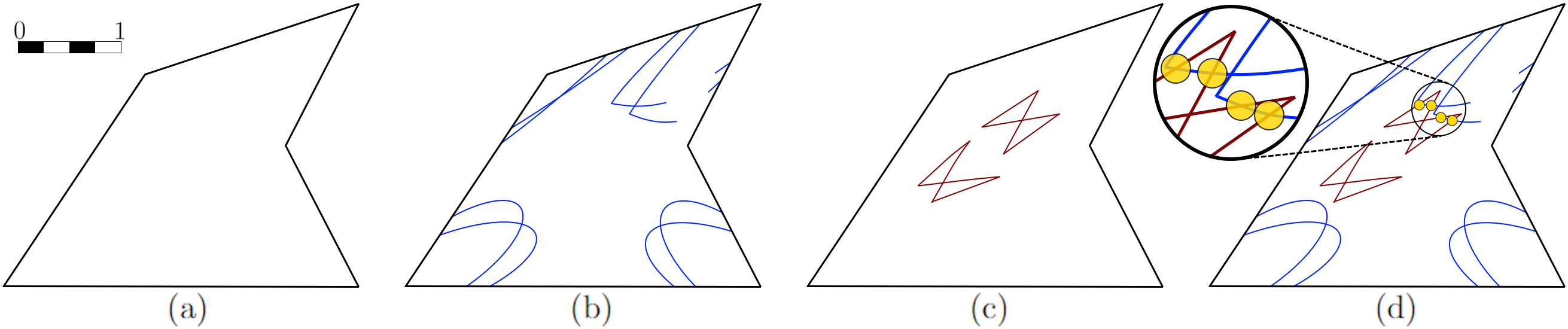}
	\caption{ \label{fig:overlay-of-antipodal} \sf
		Given a workspace (a), if a sensor measured two measurements $d_1=0.6,d_2=0.8$ in antipodal directions, its possible locations are shown in (b).
		On the other hand, if the measurements values were $d_1=1.4,d_2=2.2$, its possible locations are shown at (c).
		If we know a sensor measured the first two antipodal measurements, then rotated to some unknown direction, and then measured the second pair of antipodal measurements, the possible locations of the sensor are the intersection of (b) and (c), shown in (d).
		Assuming general position, there will be a finite set of points the sensor might be in, marked by yellow discs in (d).
	}
\end{figure}

We have presented data structures to retrieve the unknown pose of a sensor in a known polygonal environment, in an output-sensitive fashion, based on depth measurements. The data structure for a single depth measurement provides two-dimensional surface patches as answers, and the data structure for two-antipodal measurements provides elliptical arcs as answers. The obvious next goal is to add one or more measurements from the same position in the workspace to reduce the answers to a finite number of poses. Figure~\ref{fig:overlay-of-antipodal} suggests an easy way to exploit our antipodal-measurement results to
reduce the set of potential poses to a finite set of points, albeit no longer in an output-sensitive manner.

Additionally, we wish to allow movement of the sensor (robot) between measurements. Extending the exact analysis of the type we have applied in this work to more measurements or to allowing movement between measurements seems rather hard.
We remark that our analysis here has already been used in a companion paper~\cite{DBLP:conf/icra/BilevichLH23}, which takes a numerical approach to the problem,  in a system that makes three or more measurements, and using numerical methods well approximates the pose of the sensor, both in simulation and with a physical robot.

The data structures described in the paper are intended to be computed once and queried many times, but in other use cases a workspace may be queried only once.
What is the best time complexity required to answer a single query without any preprocessing?
Another variant to consider is one in which dynamic modifications of the workspace are allowed in the form of removing a vertex and its edges and inserting a new edge connecting its adjacent vertices, or adding a new vertex connected with two new edges to an existing edge endpoints as a replacement for it.
Can one construct dynamic data structures that can be updated efficiently under these modifications, and answer queries in an output-sensitive manner?
Finally a major open issue is to make the algorithm robust to imprecision in the workspace modelling or the sensor readings.

\newpage{}
\appendix
\section{Plane Projection of Measurement Results}

\subsection{Plane Projection Curves of a Single Measurement Result}
\label{app:single-projection}

Given a sensor reading $d$,
in \Cref{subsec:sinlge-result-2d}, we discussed the shape of the projection of all the configurations inside a single cell $C$ of the RTD that yield the reading $d$.
In this section we analyze the projected curved $p^{C}_l$, $p^{C}_r$, which are circular arcs or conchoids of Nicomedes; see \Cref{subsec:sinlge-result-2d} for details.
Analytical description of the arcs is trivial, therefore we will focus on the description of the conchoids.
Given a cell $C(e^C_t,e^C_b,v^C_l,v^C_r,\theta^C)$, assume, w.l.o.g., that the top and bottom edges are not vertical, and denote by $e^C_t(x)=x m_{e_t} + b_{e_t}$ the equation of the line supporting the top edge and analogously for the bottom edge $e^C_b(x)=x m_{e_b} + b_{e_b}$.

The conchoid of Nicomedes is driven from a fixed point $q$, a straight line $\ell$ and a length $d$, where for every line through $q$ that intersects $\ell$, the two points on the line which are at distance $d$ from the intersection are on the conchoid.
In our case, $q$ is the limiting vertex, the straight line is the line supporting the top edge of the cell and $d$ is the query measurement.

The classic equation for a conchoid, assuming that $q=(0,0)$ and the  line $\ell$ is $x=a$, is given by:
\[ (x-a)^2(x^2+y^2)=d^2x^2 \;. \]

For the more general case, where $q=(x_0,y_0)$ for some $x_0,y_0$ and the line $\ell$ makes angle $t$ with the positive $x$-axis and is at a distance $a$ from $q$, is given by the following:
\[ s = \sin t \;,\quad c = \cos t \;, \]
\begin{align}
	\label{eqn:rotated-conchoid-raw}
	\begin{split}
		((x-x_{0})c-(y-y_{0})s-a)^2 (((x-x_{0})c-(y-y_{0})s)^2 + \\
		((x-x_{0})s+(y-y_{0})c)^2 )=d^2 ((x-x_{0})c-(y-y_{0})s)^2
	\end{split}
\end{align}

In our case, $\ell$ supports the top edge, and its angle is defined by its slope:
\[ t = \tan^{-1} (m_{e_{t}}) ,\quad \sin \tan^{-1} (m_{e_{t}}) = \frac{m_{e_{t}}}{\sqrt{1 + m_{e_{t}}^2}} ,\quad \cos \tan^{-1} (m_{e_{t}}) = \frac{1}{\sqrt{1 + m_{e_{t}}^2}} \;. \]

By using the above identities in \Cref{eqn:rotated-conchoid-raw}, we get:
\begin{multline*}
	\left(\left(x-x_{0}\right)\frac{m_{e_{t}}}{\sqrt{1+m_{e_{t}}^2 }}-\left(y-y_{0}\right)\frac{1}{\sqrt{1+m_{e_{t}}^2 }}-a\right)^2
	\\ \quad \quad \quad \quad
	\cdot \left(\left(\left(x-x_{0}\right)\frac{m_{e_{t}}}{\sqrt{1+m_{e_{t}}^2 }}-\left(y-y_{0}\right)\frac{1}{\sqrt{1+m_{e_{t}}^2 }}\right)^2
	\right. \\ \left. \quad \quad \quad \quad
	+\left(\left(x-x_{0}\right)\frac{1}{\sqrt{1+m_{e_{t}}^2 }}+\left(y-y_{0}\right)\frac{m_{e_{t}}}{\sqrt{1+m_{e_{t}}^2 }}\right)^2 \right)=
	\\
	d^2 \left(\left(x-x_{0}\right)\frac{m_{e_{t}}}{\sqrt{1+m_{e_{t}}^2 }}-\left(y-y_{0}\right)\frac{1}{\sqrt{1+m_{e_{t}}^2 }}\right)^2 \;,
\end{multline*}
\[ \begin{split}
		\frac{1}{1+m_{e_{t}}^2 }\left(\left(x-x_{0}\right)\frac{m_{e_{t}}}{\sqrt{1+m_{e_{t}}^2 }}-\left(y-y_{0}\right)\frac{1}{\sqrt{1+m_{e_{t}}^2 }}-a\right)^2 \left(\left(m_{e_{t}}\left(x-x_{0}\right)-\left(y-y_{0}\right)\right)^2
		\right. \\ \left.
		+\left(\left(x-x_{0}\right)+m_{e_{t}}\left(y-y_{0}\right)\right)^2 \right)=\frac{1}{1+m_{e_{t}}^2 }d^2 \left(m_{e_{t}}\left(x-x_{0}\right)-\left(y-y_{0}\right)\right)^2 \;,
	\end{split} \]
\[ \begin{split}
		\left(\frac{m_{e_{t}}\left(x-x_{0}\right)-\left(y-y_{0}\right)}{\sqrt{1+m_{e_{t}}^2 }}-a\right)^2 \left(m_{e_{t}}^2 \left(x-x_{0}\right)^2 -2m_{e_{t}}\left(x-x_{0}\right)\left(y-y_{0}\right)+\left(y-y_{0}\right)^2
		\right. \\ \left.
		+\left(x-x_{0}\right)^2 +2m_{e_{t}}\left(x-x_{0}\right)\left(y-y_{0}\right)+m_{e_{t}}^2 \left(y-y_{0}\right)^2 \right)=d^2 \left(m_{e_{t}}\left(x-x_{0}\right)-\left(y-y_{0}\right)\right)^2 \;,
	\end{split} \]
\begin{multline*}
	\left(\frac{m_{e_{t}}\left(x-x_{0}\right)-\left(y-y_{0}\right)}{\sqrt{1+m_{e_{t}}^2 }}-a\right)^2 \left(\left(x-x_{0}\right)^2 +\left(y-y_{0}\right)^2 \right)\left(1+m_{e_{t}}^2 \right)=\\
	d^2 \left(m_{e_{t}}\left(x-x_{0}\right)-\left(y-y_{0}\right)\right)^2 \;,
\end{multline*}
\begin{multline*}
	\left(m_{e_{t}}\left(x-x_{0}\right)-\left(y-y_{0}\right)\pm a\sqrt{1+m_{e_{t}}^2 }\right)^2 \left(\left(x-x_{0}\right)^2 +\left(y-y_{0}\right)^2 \right)=\\
	d^2 \left(m_{e_{t}}\left(x-x_{0}\right)-\left(y-y_{0}\right)\right)^2 \;.
\end{multline*}

The term with $\pm$ sign is determined by the relation between $q$ and $\ell$. If $\ell$ is above $q$, we use the plus sign, and if $\ell$ is below $q$, we use the minus sign. In summary, the equation that represent the curve $p^{C}_l$ within a cell is (can be defined analogously for $p^{C}_r$):

\begin{multline*}
	\left(m_{e_{t}}\left(x-x_{v_{l}}\right)-\left(y-y_{v_{l}}\right)\pm a\sqrt{1+m_{e_{t}}^2 }\right)^2 \left(\left(x-x_{v_{l}}\right)^2 +\left(y-y_{v_{l}}\right)^2 \right)\\
	=d^2 \left(m_{e_{t}}\left(x-x_{v_{l}}\right)-\left(y-y_{v_{l}}\right)\right)^2 \;.
\end{multline*}

\paragraph*{The Overall Region of Potential Sensor Positions for a Fixed Reading:} Given a sensor reading $d$, we retrieve all the cells with non-empty potential poses using the sorted list of the RTD-based data structure.
For each such cell $C$ we compute the region $\Psi_d(C)$.

Alternatively, we may wish to return the (two-dimensional) union of the regions $\Psi_d(C)$, for all the cells that we retrieved. See \Cref{fig:expr-m1-result} for an illustration.
Each region $\Psi_d(C)$ has a constant descriptive complexity, and if we retrieved $k$ such regions, their union can be easily computed in time $O((k+m)\log(k+m))$, where $m$ is the number of pairs of regions that intersect. In the worst case $m=O(k^2)$.
It might be the case that special properties of the regions $\Psi_d(C)$ and their juxtaposition could be used to show that the complexity of the union is $o(k^2)$. We leave this as an open problem for further research.

\subsection{Plane Projection curve of Antipodal Measurements Result}
\label{app:antipodal-projection}

Given two measurements $d_1,d_2$, and a fixed RTD cell, all positions of a sensor that can measure $d_1$ at some angle $\theta_1$ and $d_2$ at angle $\theta_2=\theta_1 + \pi$, measuring the distance to the top and bottom edges of the cell respectively, can be calculated analytically.
Consider all the line segments of length $d_1 + d_2$ with one endpoint on the top and the other on the bottom edge.
Each possible position of the sensor can be expressed as a point placed on one of these segments, at distance $d_1$ from the endpoint that lies on the top edge.
The curve that is the union of all these points, is a Glissette~\cite{Besant:1870}, which in our special case is an ellipse.

Glissettes are the curves traced out by a point on a curve $\gamma$, where $\gamma$ slides along two other fixed curves.
In our case specifically, $\gamma$ is a fixed line segment $s$, sliding while each of its endpoints is in contact with one line (the two lines are not necessary orthogonal to one another), and the Glissette is defined as the trace of an arbitrary point on $s$ (not necessarily the mid-point).
A solution for the case where the two fixed lines are at right angle is given at \cite[p.~51]{Besant:1870}.
We solve it here for the general case.

For simplicity, assume that one of the fixed lines is the $x$-axis, it intersect the other fixed line at the origin, and the angle between them is $\alpha$.
If the moving line segment has length $d_1+d_2$, and the designated point to be traced is at distance $d_1,d_2$ from the segment's endpoints, the Glissette points satisfy the following, for any angle $\phi$ in \textbf{oblique} coordinates:
\[ x \sin\alpha = f(\phi) \;,\quad y \sin\alpha = f(\phi +\alpha) \;. \]

Where $f(\cdot)$ is the \textbf{tangential polar equation}.
In our case:
\[ f(\phi) =\begin{cases}
		d_1 \cos\phi, & \text{ if } \phi \in [0, \pi/2], \\
		d_2 \cos\phi, & \text{ else }
	\end{cases} \;. \]

Again, these expressions represent the oblique coordinates.
When transforming into Cartesian coordinates, we get:
\[ y=d_2 \cos(\phi+\alpha) \;,\quad x=\frac{d_1 \cos\phi}{\sin\alpha}+\frac{y}{\tan\alpha} \;. \]

The above expressions are correct for any angle $\phi$, and we would like to express all the result points in a single equation.
Next, we perform angle elimination:
\[ \theta = \arccos (\frac{y}{d_2}) -\alpha \;, \]
\[ x = \frac{d_1 \cos(\arccos (\frac{y}{d_2}) -\alpha)}{\sin\alpha} + \frac{y}{\tan \alpha} \;, \]
\[ x = \frac{d_1 (\cos\arccos\frac{y}{d_2}\cos\alpha+\sin\arccos\frac{y}{d_2}\sin\alpha)}{\sin\alpha} + \frac{y}{\tan \alpha} \;, \]
\[ x = \frac{d_1 (\frac{y}{d_2}\cos\alpha \pm\sqrt{1-\frac{y^2}{d_2^2}}\sin\alpha)}{\sin\alpha} + \frac{y}{\tan \alpha} \;, \]
\[ x = d_1 (\frac{y}{d_2}\frac{\cos\alpha}{\sin\alpha} \pm\sqrt{1-\frac{y^2}{d_2^2}}) + \frac{y}{\tan \alpha} \;, \]
\[ x = \frac{d_1}{d_2}\frac{y}{\tan\alpha} \pm d_1 \sqrt{1-\frac{y^2}{d_2^2}} + \frac{y}{\tan \alpha} \;, \]
\[ x -(1+\frac{d_1}{d_2})\frac{y}{\tan\alpha} = \pm d_1 \sqrt{1-\frac{y^2}{d_2^2}} \;, \]
\[ x^2 -2x(1+\frac{d_1}{d_2})\frac{y}{\tan\alpha} +(1+\frac{d_1}{d_2})^2\frac{y^2}{\tan^2\alpha} = d_1^2 (1-\frac{y^2}{d_2^2}) \;, \]
\[ x^2 -2\frac{1+\frac{d_1}{d_2}}{\tan\alpha}xy +\left(\frac{(1+\frac{d_1}{d_2})^2}{\tan^2\alpha} +\frac{d_1^2}{d_2^2}\right) y^2 =d_1^2 \;, \]
\begin{equation} \label{eqn:ellipse_raw}
	\frac{x^2}{d_1^2} -2\frac{1+\frac{d_1}{d_2}}{d_1^2\tan\alpha}xy +\left(\frac{(1+\frac{d_1}{d_2})^2}{d_1^2\tan^2\alpha} +\frac{1}{d_2^2}\right) y^2 =1 \;.
\end{equation}

The above equation already expresses all the points in the ellipse, but we would like to work with standard representation of rotated ellipses.
Rotated ellipse equation centered at $(x_0,y_0)$ with rotation angle $\theta$:
\[ \frac{((x-x_0)\cos\theta +(y-y_0)\sin\theta)^2}{a^2} +\frac{((x-x_0)\sin\theta -(y-y_0)\cos\theta)^2}{b^2} = 1 \;. \]
We assumed that the intersection is at the origin, therefore $x_0=y_0=0$.
Denote $\sin\theta=s, \cos\theta=c, \tan\theta=t$, then
\[ \frac{(xc +ys)^2}{a^2} +\frac{(xs -yc)^2}{b^2} = 1 \;, \]
\[ (\frac{c^2}{a^2} +\frac{s^2}{b^2})x^2 +2cs(\frac{1}{a^2} -\frac{1}{b^2})xy + (\frac{s^2}{a^2} +\frac{c^2}{b^2})y^2 =1 \;. \]
From \Cref{eqn:ellipse_raw} and the above we can derive three equations with three unknowns:
\begin{equation} \label{eqn:ellipse_x2} \frac{1}{d_1^2} = \frac{c^2}{a^2} +\frac{s^2}{b^2} \;, \end{equation}
\begin{equation} \label{eqn:ellipse_xy} -2\frac{1+\frac{d_1}{d_2}}{d_1^2\tan\alpha} = 2cs(\frac{1}{a^2} -\frac{1}{b^2}) \;, \end{equation}
\begin{equation} \label{eqn:ellipse_y2} \frac{(1+\frac{d_1}{d_2})^2}{d_1^2\tan^2\alpha} +\frac{1}{d_2^2} =\frac{s^2}{a^2} +\frac{c^2}{b^2} \;. \end{equation}

We can solve these equations to calculate $a,b,\theta$. \\
From \Cref{eqn:ellipse_x2} it follows that
\[ \frac{1}{d_1^2} = \frac{c^2}{a^2} +\frac{s^2}{b^2} =\frac{c^2b^2 +s^2a^2}{a^2b^2} \;, \]
\[ a^2b^2 =d_1^2(c^2b^2 +s^2a^2) \;, \]
\begin{equation} \label{eqn:ellipse_a2_dirty} a^2 =\frac{d_1^2 c^2b^2}{b^2 -d_1^2 s^2} \;. \end{equation}
From \Cref{eqn:ellipse_xy} it follows that
\[ -2\frac{1+\frac{d_1}{d_2}}{d_1^2\tan\alpha} = 2cs(\frac{1}{a^2} -\frac{1}{b^2}) \;, \]
\[ \frac{1+\frac{d_1}{d_2}}{csd_1^2\tan\alpha} = \frac{1}{b^2} -\frac{1}{a^2} \;. \]
Assigning $a^2$ using \Cref{eqn:ellipse_a2_dirty}, we get
\[ \frac{1+\frac{d_1}{d_2}}{csd_1^2\tan\alpha} = \frac{1}{b^2} -\frac{b^2 -d_1^2 s^2}{d_1^2 c^2b^2} \;, \]
\[ \frac{1+\frac{d_1}{d_2}}{csd_1^2\tan\alpha} = \frac{d_1^2 c^2 - b^2 +d_1^2 s^2}{d_1^2 c^2b^2} \;, \]
\[ \frac{1+\frac{d_1}{d_2}}{s \tan\alpha} = \frac{d_1^2 - b^2}{cb^2} \;, \]
\[ \frac{c(1+\frac{d_1}{d_2})}{s \tan\alpha} +1 = \frac{d_1^2}{b^2} \;, \]
\begin{equation} \label{eqn:ellipse_b2} b^2 =\frac{d_1^2}{\frac{c(1+\frac{d_1}{d_2})}{s \tan\alpha} +1} =\frac{d_1^2}{1 +\frac{c}{s \tan\alpha}(1 +\frac{d_1}{d_2})} \;. \end{equation}
Assigning $b^2$ using \Cref{eqn:ellipse_b2} into \Cref{eqn:ellipse_a2_dirty}, we get:
\[ a^2 =\frac{d_1^2 c^2b^2}{b^2 -d_1^2 s^2} =\frac{d_1^2 c^2}{1 -\frac{d_1^2 s^2}{b^2}} =\frac{d_1^2 c^2}{1 -\frac{d_1^2 s^2}{\frac{d_1^2}{1 +\frac{c}{s \tan\alpha}(1 +\frac{d_1}{d_2})}}} \;, \]
\[ =\frac{d_1^2 c^2}{1 -s^2 -\frac{cs}{\tan\alpha}(1 +\frac{d_1}{d_2})} =\frac{d_1^2 c^2}{c^2 -\frac{cs}{\tan\alpha}(1 +\frac{d_1}{d_2})} =\frac{d_1^2}{1 -\frac{s}{c\tan\alpha}(1 +\frac{d_1}{d_2})} \;, \]
\begin{equation} \label{eqn:ellipse_a2} a^2 =\frac{d_1^2}{1 -\frac{s}{c\tan\alpha}(1 +\frac{d_1}{d_2})} \;. \end{equation}

We can now return to \Cref{eqn:ellipse_xy}, and replace $a^2$ and $b^2$ with the expressions obtained in \Cref{eqn:ellipse_a2} and \Cref{eqn:ellipse_b2} respectively:

\[ \frac{(1+\frac{d_1}{d_2})^2}{d_1^2\tan^2\alpha} +\frac{1}{d_2^2} =\frac{s^2}{a^2} +\frac{c^2}{b^2} \;, \]
\[ \frac{(1+\frac{d_1}{d_2})^2}{d_1^2\tan^2\alpha} +\frac{1}{d_2^2} =\frac{s^2}{\frac{d_1^2}{1 -\frac{s}{c\tan\alpha}(1 +\frac{d_1}{d_2})}} +\frac{c^2}{\frac{d_1^2}{1 +\frac{c}{s \tan\alpha}(1 +\frac{d_1}{d_2})}} \;, \]
\[ \frac{(1+\frac{d_1}{d_2})^2}{\tan^2\alpha} +\frac{d_1^2}{d_2^2} =s^2(1 -\frac{s}{c\tan\alpha}(1 +\frac{d_1}{d_2})) +c^2(1 +\frac{c}{s \tan\alpha}(1 +\frac{d_1}{d_2})) \;, \]
\[ \frac{(1+\frac{d_1}{d_2})^2}{\tan^2\alpha} +\frac{d_1^2}{d_2^2} =s^2 -\frac{ts^2}{\tan\alpha}(1 +\frac{d_1}{d_2}) +c^2 +\frac{c^2}{t \tan\alpha}(1 +\frac{d_1}{d_2}) \;, \]
\[ \frac{(1+\frac{d_1}{d_2})^2}{\tan^2\alpha} +\frac{d_1^2}{d_2^2} =1 +\frac{c^2 -t^2s^2}{t\tan\alpha}(1 +\frac{d_1}{d_2}) \;, \]
\[ \frac{(1+\frac{d_1}{d_2})^2}{\tan^2\alpha} +\frac{d_1^2}{d_2^2} -1 =\frac{1-2s^2}{tc^2\tan\alpha}(1 +\frac{d_1}{d_2}) \;, \]
\[ \frac{(1+\frac{d_1}{d_2})^2}{\tan^2\alpha} +(\frac{d_1}{d_2} -1)(\frac{d_1}{d_2} +1) =\frac{1-2s^2}{tc^2\tan\alpha}(1 +\frac{d_1}{d_2}) \;, \]
\[ \frac{(1+\frac{d_1}{d_2})}{\tan^2\alpha} +\frac{d_1}{d_2} -1 =\frac{1-2s^2}{tc^2\tan\alpha} \;, \]
\[ \frac{(1+\frac{d_1}{d_2})}{\tan\alpha} +\frac{d_1 \tan\alpha}{d_2} -\tan\alpha =\frac{1-2s^2}{tc^2} \;, \]
\[ \frac{d_1+d_2 +d_1 \tan^2\alpha -d_2\tan^2\alpha}{d_2\tan\alpha} =\frac{1-2s^2}{tc^2} \;, \]
\[ \frac{d_1(1 +\tan^2\alpha) +d_2(1 -\tan^2\alpha)}{d_2\tan\alpha} =\frac{1-2s^2}{tc^2} \;, \]
\[ \frac{1-2s^2}{c^2t} = \frac{cos(2\theta)}{c^2t} =2\cot(2\theta) \;, \]
\[ 1 +\tan^2\alpha = \frac{1}{\cos^2 \alpha} ,\quad 1 -\tan^2\alpha =\frac{1-2\sin^2\alpha}{\cos^2\alpha} =\frac{\cos(2\alpha)}{\cos^2\alpha} \;, \]
\[ \frac{\frac{d_1}{\cos^2\alpha} +\frac{d_2\cos(2\alpha)}{\cos^2\alpha}}{d_2\tan\alpha} =2\cot(2\theta) \;, \]
\[ \frac{d_1 +d_2\cos(2\alpha)}{d_2 \sin\alpha \cos\alpha} =2\cot(2\theta) \;, \]
\[ \frac{\frac{d_1}{d_2} +\cos(2\alpha)}{\sin(2\alpha)} =\cot(2\theta) \;, \]
\begin{equation} \label{eqn:ellipse_theta} \theta = \frac{1}{2}\cot^{-1}\left(\frac{\frac{d_1}{d_2} +\cos(2\alpha)}{\sin(2\alpha)}\right) \;. \end{equation}

With $\theta$ calculated, we can go back to \Cref{eqn:ellipse_a2} and \Cref{eqn:ellipse_b2} and easily calculate their exact value.
We have calculate the fixed values $a,b,\theta$ under the assumption that the line supporting the bottom edge is the $x$-axis and it intersects the line supporting the top edge at the origin.
The ellipse is expressed by the equation:

\[ \frac{(x \cos\theta +y \sin\theta)^2}{a^2} +\frac{(x \sin\theta -y \cos\theta)^2}{b^2} = 1 \;. \]

In full generality, for any cell, we use $\theta' = \theta + \tan^{-1}(m^C_b)$, and offset the center of the ellipse to the intersection of the lines supporting the top and bottom edge of the cell $(x_0, y_0)$:

\[ \frac{((x-x_0)\cos\theta' +(y-y_0)\sin\theta')^2}{a^2} +\frac{((x-x_0)\sin\theta' -(y-y_0)\cos\theta')^2}{b^2} = 1\;. \]

\section{Combinatorial Result Change Handling}
\label{app:comb-change-ops}

In this section, we provide more details on the analysis in \Cref{subsec:Combinatorial-result-change-handling}.
All the tuples with the same first parameter $e$, sorted in increasing $d$ value, are the events of the spatial sweep with the surface $\Sigma(e,d)$.
We compute the result $\sigma(e,d)$ for infinitesimal values of $d$, by considering only $e$ and its two adjacent edges.
From this points on, result edges will be inserted or removed only when handling combinatorial change tuples.
We handle each combinatorial change $(e, \theta, d, U)$ according to the event type.
We fully describe how to handle a change $(e,\theta,d,U)$ of type I and V, as types II, III and IV are analog to I.
Let us first introduce some definitions and general notes:

\begin{enumerate}[(a)]
	\item
	      We say that a result edge $\eta'$ is defined by $e$ and a vertex $v$ when $\eta'$ is the union of configurations where a rod of length $d$ is sliding on $e$ and $v$, and we denote this rod by $s(e,v)$.

	\item
	      We say that a result edge $\eta''$ is defined by $e$ and another edge $e_1$ when $\eta''$ is the union of configurations where a rod of length $d$ is sliding with one endpoint on $e$ and another endpoint on $e_1$, and we denote this rod by $s(e,e_1)$.

	\item
	      The limits $\alpha, \beta$ of the angle range $[\alpha, \beta]$ throughout which a result edge prevails, can be specified by a fixed angle each, or by the labels of the workspace features that induced these angles for a specific $d$ value.

	\item
	      Note that we refer to 'the' segment (or 'the' rod) of length $d$ with its endpoints on some vertex and $e$, while there may be two such segments.
	      The uniqueness of the segment is specified by the constraint on its angle, for example $\delta \leq \theta$ in the first sub-case of type I handling.

	\item
	      For the simplicity of exposition, we assume that $e$ is parallel to the $x$-axis.

	\item
	      We say that a vertex is a \textit{reflex vertex relative to a line (resp.\ segment)} if both of its incident edges are in the same half-plane induced by the line (resp.\ the line supporting the segment).

	\item
	      By 'the angle of a segment/rod' we refer to the angle that the line supporting the segment/rod creates with the positive direction of the $x$-axis.

\end{enumerate}

\begin{description}
	\item[Combinatorial change of type~\textbf{\textnormal{I}}]:
	      $d$ is the minimum value in which a result vertex of type~(i)
	      involving two vertices $v_1,v_2$ exists (namely an intersection of two $\chi_e(\cdot)$ surfaces intersect $\Sigma(e,d)$), where $v_1$ is the vertex closer to $e$.
	      We use the following assumptions and definitions:
	      \begin{enumerate}[(a)]
		      \item
		            We assume that $v_2$ is a reflex vertex relative to $\overline{v_1v_2}$; otherwise, $d$ is also the maximum value in which such a result vertex exists, and we will handle it as a change of type~II (see \Cref{fig:result-vertices-comb-changes-sweep}(a) for such a scenario).

		      \item
		            The vertex $v_1$ is clearly a reflex vertex relative to $\overline{v_1v_2}$, and we will assume  that the edges incident to $v_2$ are on the left halfplane relative to the direction $\overrightarrow{v_2v_1}$.

		      \item
		            Let $e_1$ be the edge of $v_2$ visible from $v_1$.
	      \end{enumerate}

	      The handling operations depend on two predicates:
	      whether the edges of both $v_1,v_2$ are in the same halfplane relative to $\overline{v_1v_2}$ or not, and whether a result edge defined by $e,e_1$ exists or not in $\sigma(e,d')$ for $d'$ infinitesimally smaller than $d$.
	      The sub-cases are illustrated in \Cref{fig:result-vertices-comb-change-I-subcases}.

	      \begin{itemize}
		      \item
		            A result edge $\eta_1$ defined by $e,e_1$ exists in the angle range $[\alpha, \gamma]$ for some $\alpha$, where $\gamma$ depends on $d$ and it is defined as the angle that $s(e,e_1)$ (that also passes through $v_1$ at the time of handling the event) makes with the $x$-axis.
		            There must be another result edge $\eta_2$ defined by $e,v_1$ in the angle range $[\gamma, \beta]$ for some $\beta$.
		            \begin{itemize}
			            \item
			                  If the edges of both $v_1,v_2$ are in the same halfplane relative to $\overline{v_1v_2}$:
			                  \begin{itemize}
				                  \item
				                        Find and remove $\eta_1$ and $\eta_2$.
				                  \item
				                        Add a result edge defined by $e,v_2$ in the angle range $[\delta, \theta]$ where $\delta \leq \theta$ is the angle of the rod $s(e,v_2)$ (there are two such rods and we choose one of them by the constraint $\delta \leq \theta$).
				                        Recall that $\theta$ is a parameter of the combinatorial change tuple.
				                  \item
				                        Add a result edge defined by $e,v_1$ in the angle range $[\theta, \beta]$.
			                  \end{itemize}

			            \item
			                  If the edges of $v_1,v_2$ are not in the same halfplane relative to $\overline{v_1v_2}$:
			                  \begin{itemize}
				                  \item
				                        Find and remove $\eta_1$ and $\eta_2$.
				                  \item
				                        Add a result edge defined by $e,v_2$ in the angle range $[\theta, \delta]$ where $\delta \geq \theta$ is the angle of the rod $s(e,v_2)$.
				                  \item
				                        Add a result edge defined by $e, v_1$ in the angle range $[\theta, \beta]$.
			                  \end{itemize}
		            \end{itemize}

		      \item
		            A result edge defined by $e,e_1$ does not exists.
		            Therefore, there must be a result edge $\eta_3$ defined in the angle range $[\alpha, \beta]$ where $\theta \in (\alpha, \beta)$.
		            \begin{itemize}
			            \item
			                  If the edges of both $v_1,v_2$ are in the same halfplane relative to $\overline{v_1v_2}$:
			                  \begin{itemize}
				                  \item
				                        Find and remove $\eta_3$.
				                  \item
				                        Add a result edge defined by $e,e_1$ in the angle range $[\alpha, \phi]$ where $\phi \geq \alpha$ is the angle of the rod $s(e,e_1)$ (that also passes through $v_1$ at the time of handling the event).
				                  \item
				                        Add a result edge defined by $e,v_2$ in the angle range $[\delta, \theta]$ where $\delta \leq \theta$ is the angle of the rod $s(e<v_2)$.
				                  \item
				                        Add a result edge defined by $e,v_1$ in the angle range $[\theta, \beta]$.
			                  \end{itemize}
		            \end{itemize}

		      \item
		            If the edges of $v_1,v_2$ are not in the same halfplane relative to $\overline{v_1v_2}$:
		            \begin{itemize}
			            \item
			                  Find and remove $\eta_3$.
			            \item
			                  Add a result edge defined by $e,e_1$ in the angle range $[\alpha, \phi]$ where $\phi \geq \alpha$ is the angle of the rod $s(e,e1)$ (that also passes through $v_1$ at the time of handling the event).
			            \item
			                  Add a result edge defined by $e,v_2$ in the angle range $[\theta, \delta]$ where $\delta \geq \theta$ is the angle of the rod $s(e,v_2)$.
			            \item
			                  Add a result edge defined by $e,v_1$ in the angle range $[\theta, \beta]$.
		            \end{itemize}
	      \end{itemize}

	\item[Combinatorial change of type~\textbf{\textnormal{V}}]:
	      $d$ is the length of the shortest line segment, denoted $\bar{s}$, which is contained in the workspace, from a vertex $v$ to $e$.
	      We use the following assumptions and definitions:
	      \begin{enumerate}[(a)]
		      \item
		            Let $e_1,e_2$ be the edges incident to~$v$, where $e_1$ is
		            met first when we go counterclockwise around the vertex $v$ from $\bar{s}$.

		      \item
		            Let $\theta$ be the angle create by the segment $\bar{s}$ and the positive $x$-axis.

		      \item
		            We assume that $e_2$ is in the right halfplane relative to $\bar{s}$, where $\bar{s}$ is directed from $v$ to its endpoint on $e$.
		            The case where $e_2$ is in the left halfplane can be handled symmetrically.

		      \item
		            Let $\gamma \leq \theta$ and $\phi \geq \theta$ be the angles of the two rods $s(e,v)$ (while typically we refer to a single rod, as we have some constraint on the rod orientation, in this case we consider both rods.
	      \end{enumerate}
	      The handling operations of the change depend on whether there are already result edges defined by $e_1$ and $e_2$:

	      \begin{itemize}
		      \item
		            There are result edges defined by $e,e_1$ and $e,e_2$ (\Cref{fig:result-vertices-comb-change-V-subcases}(a)), denoted by $\eta_1$ and $\eta_2$ respectively:
		            \newline Let $[\alpha_1, \beta_1], [\alpha_2, \beta_2]$ be the angle ranges $\eta_1, \eta_2$ exist in, respectively.
		            \begin{itemize}
			            \item
			                  Find and remove $\eta_1$ and $\eta_2$.
			            \item
			                  Add a result edge defined by $e,e_1$ in the angle range $[\alpha_1, \gamma]$.
			            \item
			                  Add a result edge defined by $e,e_1$ in the angle range $[\phi, \beta_1]$.
			            \item
			                  Add a result edge defined by $e,e_2$ in the angle range $[\alpha_2, \gamma]$.
			            \item
			                  Add a result edge defined by $e,e_2$ in the angle range $[\phi, \beta_2]$.
		            \end{itemize}

		      \item
		            There is a single result edge defined by either $e,e_1$ or $e,e_2$:
		            \newline Assume that only the result edge defined by $e,e_2$ exists, and denote it by $\eta_2$. The case where only the result edge defined by $e,e_1$ exists is handled symmetrically.
		            First, $\eta_2$ should be removed from $R$.
		            The result edges that should be added depends on these sub cases:
		            \begin{itemize}
			            \item
			                  Both $e_1,e_2$ are in the right halfplane relative to $\bar{s}$ (\Cref{fig:result-vertices-comb-change-V-subcases}(b)):
			                  \begin{itemize}
				                  \item
				                        A result edge defined by $e,v$ that exists in the angle range $[\gamma, \phi]$.
				                  \item
				                        Add a result edge defined by $e,e_2$ in the angle range $[\alpha_2, \gamma]$.
				                  \item
				                        Add a result edge defined by $e,e_2$ in the angle range $[\phi, \beta_2]$.
			                  \end{itemize}
			            \item
			                  The edges $e_1, e_2$ are in different halfplanes relative to $\bar{s}$ (\Cref{fig:result-vertices-comb-change-V-subcases}(c)):
			                  \begin{itemize}
				                  \item
				                        A result edge defined by $e,e_1$ that exists in the angle range $[\gamma, \phi]$.
				                  \item
				                        Add a result edge defined by $e,e_2$ in the angle range $[\alpha_2, \gamma]$.
				                  \item
				                        Add a result edge defined by $e,e_2$ in the angle range $[\phi, \beta_2]$.
			                  \end{itemize}
		            \end{itemize}

		      \item
		            There are no result edges defined by $e,e_1$ or $e,e_2$:
		            \newline A new hole in the result should be added, with edges depending on the following sub-cases:
		            \begin{itemize}
			            \item
			                  Both $e_1,e_2$ are in the right halfplane relative to $\bar{s}$ (\Cref{fig:result-vertices-comb-change-V-subcases}(d)):
			                  \begin{itemize}
				                  \item
				                        A result edge defined by $e,v$ that exists in the angle range $[\gamma, \phi]$.
				                  \item
				                        A result edge defined by $e,e_2$ that exists in the angle range $[\gamma, \phi]$.
			                  \end{itemize}
			            \item
			                  The edges $e_1, e_2$ are in different halfplanes relative to $\bar{s}$ (\Cref{fig:result-vertices-comb-change-V-subcases}(e)):
			                  \begin{itemize}
				                  \item
				                        A result edge defined by $e,e_1$ that exists in the angle range $[\gamma, \phi]$.
				                  \item
				                        A result edge defined by $e,e_2$ that exists in the angle range $[\gamma, \phi]$.
			                  \end{itemize}
		            \end{itemize}
	      \end{itemize}

\end{description}

\begin{figure}[h] \centering
	\includegraphics[page=1, scale=0.85]{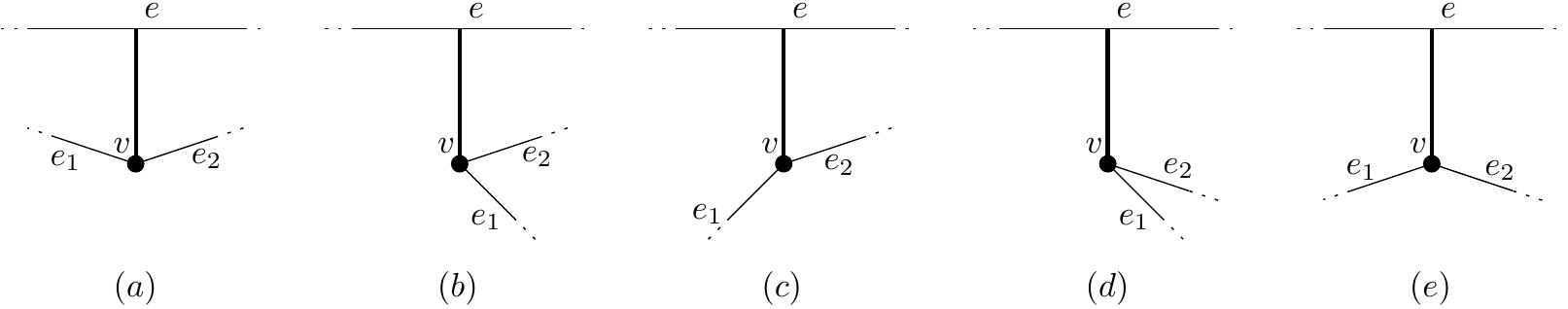}
	\caption{ \label{fig:result-vertices-comb-change-V-subcases} \sf
		Examples of the sub-cases of handling combinatorial change of type (V).
		The bold line is a segment of length $d$.
		There is a result edge defined by $e,e_1$ fof length $<d$ in (a).
		There is a result edge defined by $e,e_2$ of length $<d$ in (a,b,c).
		Both $e_1$ and $e_2$ are in the same halfplane relative to the shortest line between $v$ and $e$, in (b)(d).
	}
\end{figure}

\bibliographystyle{abbrv}
\bibliography{references.bib}

\end{document}